# “Aftereffects” Phenomenon in $^{111}In(\rightarrow^{111}Cd)$-Implanted $\alpha$-$Al_2O_3$ Single Crystals: Novel Approach Integrating Experimental Double-Model Analysis with Density-Functional Theory

*Germán N. Darriba* [*,†], *Reiner Vianden* [‡], *Alejandro P. Ayala* [§], *and Mario Rentería* [*,†]

† Departamento de Física “Prof. Dr. Emil Bose” and Instituto de Física La Plata [IFLP, Consejo Nacional de Investigaciones Científicas y Técnicas (CONICET) La Plata], Facultad de Ciencias Exactas, Universidad Nacional de La Plata, CC 67, (1900) La Plata, Argentina.

‡ Helmholtz-Institut für Strahlen- und Kernphysik (HISKP), Universität Bonn, Nussallee 14–16, 53115 Bonn, Germany

§ Departamento de Física, Universidade Federal do Ceará, Fortaleza, CE, 60644-900, Brazil.

ABSTRACT: We develop an experimental double-model analysis, combined with density-functional theory (DFT), to explore the origins of dynamic hyperfine interactions (HFIs) linked to the electron-capture decay "aftereffects" (ECAE) phenomenon. This electronic effect, reversible with temperature, has been observed in time-differential perturbed γ-γ angular correlations (TDPAC) experiments on oxides doped with ($^{111}$In (EC)→)$^{111}$Cd probe atoms. Besides identifying the electronic configuration that yields the stable final electric-field gradient (EFG) after the dynamic process ends, we determine the *initial* configurations around the probe nucleus and their corresponding EFGs whose fluctuations produce these dynamic HFIs. We demonstrate the equivalence between parameters of the two most widely used methods for analyzing this type of dynamic HFI, enabling us to obtain these initial electronic configurations at *each* temperature. In this framework, to unravel controversial TDPAC results reported for $^{111}$In-implanted α-$Al_2O_3$ single crystals, we perform a DFT study of Cd-doped α-$Al_2O_3$, examining their defect-formation energies, as functions of the Cd impurity level´s charge state. We show that the stable final EFG for the expected interaction $HFI_u$ originates from $^{111}$Cd probes located at defect-free substitutional Al sites (without trapped electron holes) across all measured temperatures. Those of the unexpected $HFI_d$ originate from probes at Al sites, but with different degrees of occupation of the Cd impurity level. The correct analysis of $HFI_d$ provides experimental evidence quantifying the charge states of the $^{111}$Cd atom during its electronic recovery. The electron holes´ lifetime $\tau_g$ (ranging from 5 to 45 ns) provides information about the system's electron availability and mobility. We show that one trapped hole for $HFI_u$ and at least five for $HFI_d$ are responsible for the dynamic regime when the "aftereffects" are more pronounced. The proposed scenario accounts for the observation of well-defined EFGs when the dynamic regime does not end.

.

## I. INTRODUCTION

The incorporation of atomic impurities into host lattices can lead to the emergence of new physical properties, as observed, e.g., in semiconducting oxides.[1-5] Experimental techniques, such as nuclear solid-state methods that often use impurity atoms as probes, are essential for investigating local effects at the atomic and sub-nanoscopic levels. A powerful tool for studying subtle changes in the electronic structure near probe atoms is the time-differential perturbed γ-γ angular correlations (TDPAC) technique, with the ($^{111}$In→)$^{111}$Cd nuclide being the most used probe atom. This nuclear spectroscopy measures, in particular, the hyperfine interaction (HFI) between the nuclear electric quadrupole moment $Q$ of the probe nucleus and the extranuclear electric-field gradient (EFG) tensor $V_{ij}(\vec{r}) = \partial^2 V(\vec{r}) / \partial x_i \partial x_j$ at the probe site, being $V(\vec{r})$ the Coulomb potential. If a fluctuating EFG (*dynamic* HFI) is sensed by the probe atom rather than the usual *static* EFG, the $R(t)$ TDPAC spectrum exhibits a characteristic strong damping.

In many TDPAC experiments studying $^{111}$In(→$^{111}$Cd)-doped binary oxides, dynamic HFIs have been observed.[6-16] The electron-capture (EC) decay of the $^{111}$In isotope into $^{111}$Cd generates in the probe an outward cascade of multiple electron holes through successive Auger processes. If the charge recombination of the last electron holes of the outermost shells occurs during the time window of the experimental TDPAC measurement (i.e., between the emission of $\gamma_1$ and $\gamma_2$ of the sensitive γ-γ cascade of the probe nucleus), a specific type of dynamic HFIs can be detected, a phenomenon known as electron-capture decay *aftereffects* (ECAE). A dynamic regime occurs when charge fluctuations around the probe nucleus induce rapid changes in the EFG magnitude, asymmetry, and/or orientation. The ($^{111}$In(EC)→)$^{111}$Cd probe (in a semiconducting or insulating environment) can generate this type of dynamic HFIs on its own, without the need to introduce impurity acceptor or donor levels in the host, as was previously thought.[16] Experimentally, the $R(t)$ spectrum exhibits a characteristic strong damping within the first nanoseconds, followed by a constant reduced spectrum´s amplitude, indicating the end (*switch off*) of the EFG fluctuation (see the naive description of Figure 2 in ref. 15). This damping decreases reversibly as the measurement temperature (T) increases, completely restoring the amplitude at higher temperatures. Contrarily, if the EFG fluctuation ends before the emission of $\gamma_1$ (indicating very fast recombination), a static HFI is observed.

To analyze this type of spectra, the *on-off* perturbation factor proposed by Bäverstam *et al.*[17] (BO approach) and the *unidirectional* electronic relaxation described by Lupascu *et al.*[11] (L approach) are the most commonly used methods in the literature. In the BO approach,[17] the perturbation factor has a straightforward analytical form, which allows the determination, at *each* temperature, of parameters that characterize the dynamic regime, along with the standard hyperfine quantities related to the *final stable* $EFG_f$ tensor. However, this perturbation factor cannot explicitly reveal the initial oscillatory EFGs that generate the dynamic behavior. On the other hand, in the L approach,[11] numerical solutions for the perturbation factor are obtained. Within this approach, it is possible to estimate the range of initial fluctuating EFGs that relax *unidirectionally* toward the final $EFG_f$, but are representative of the *entire* temperature range.[18] Unlike the BO approach, the L approach does not allow fitting a dynamic perturbation factor to the experimental data.

Combining the BO approach and *ab initio* EFG calculations as a function of the charge state of the impurity, we have proposed a scenario compatible with the "aftereffects" phenomenon in which it is possible to identify the stable electronic configurations that produce the *final* stable EFGs.[15, 16] This model proposes that rapid random fluctuations among these charge states create fast fluctuating initial EFGs, driving the system into a dynamic regime until it decays to a final stable electronic configuration that produces the $EFG_f$ observed in this type of TDPAC spectra. If the initial oscillating EFGs could be determined at *each* temperature using the methodology proposed in this paper, *ab initio* EFG calculations would identify which electronic configurations give rise to the dynamic regime.

Penner *et al.*[14] reported dynamic hyperfine interactions in $^{111}$In(EC→$^{111}$Cd)-doped $\alpha$-$Al_2O_3$ single crystals exhibiting intriguing and controversial temperature-dependent behavior (4-973 K). They observed two interactions: $HFI_u$ (*undisturbed*) and $HFI_d$ (*disturbed*). $HFI_u$ denotes $^{111}$Cd atoms occupying single-crystalline Al sites free of structural defects. Conversely, $HFI_d$ was assigned to $^{111}$Cd atoms occupying Al sites, with randomly distributed structural defects around the probes. Only $HFI_u$ was analyzed for dynamic behavior, within the L approach. Contrary, $HFI_d$ was artificially treated as a *static polycrystalline* interaction with a fictitious wide (static) EFG distribution, although the Cd atoms are located at *single-crystalline* sites. To resolve this controversy, we revisit the experiments using an integrated BO and L approach, along with a comprehensive *ab initio* study of the doped system, thereby obtaining all the information contained in the spectra and clarifying the origin of the "aftereffects" phenomenon.

To achieve these goals, it was first essential to demonstrate, under certain assumptions, the equivalence of the two approaches, relating their dynamic parameters associated with the range of initial EFGs. Then, by applying the BO model to the experiment, we can determine the initial fluctuating EFGs that drive the dynamics for *both* interactions, at *each* measurement temperature. The unexpected presence of $HFI_d$ in the present work serves as an excellent "laboratory" to demonstrate the existence of different charge states of the $^{111}$Cd probe during the latter stages of the electronic recovery process.

To predict the final stable EFGs observed experimentally and the initial fluctuating EFGs determined using the double-model approach for both interactions, we perform a thorough *ab initio* study of diluted Cd-doped $\alpha$-$Al_2O_3$ as a function of the system´s charge state. In addition, we calculate defect formation energies for all charged systems to evaluate their stability as final states and the likelihood of fluctuations between specific initial charge states. An analysis of these theoretical results, combined with the integrated experimental analysis, enables us to quantify the electronic configurations (associated with trapped electron holes) that give rise to the initial fluctuating EFGs and the stable final EFGs.

In summary, we introduce a synergistic approach, supported by *ab initio* calculations, that enables analysis of any spectrum exhibiting dynamic hyperfine interactions. This method allows extraction of the $EFG_f$, the set of initial fluctuating EFGs, and the different electronic configurations that give rise to them, specifically the number of electron holes and their lifetimes. These details offer valuable insights into the electrical properties, such as electron availability and mobility, within the host-impurity system. Additionally, when multiple interactions are present, this approach helps determine whether each interaction is dynamic or static.

## II. PERTURBATION FACTORS FOR DYNAMIC HYPERFINE INTERACTIONS

TDPAC spectroscopy enables precise measurement of hyperfine interactions between external electric and magnetic fields and the nuclear quadrupole and/or dipole moments of the intermediate state in a suitable $\gamma$-$\gamma$ cascade decay.[19-21] This is the case of the well-known 171-245 keV $\gamma_1$-$\gamma_2$ cascade in the $^{111}$Cd probe nucleus, which results from the EC decay of the $^{111}$In parent. Experimentally, the $R$(t) spectrum is built by measuring the number of events (coincidences) where $\gamma_2$ is detected after a time $t$ from the $\gamma_1$ detection (the time window of the TDPAC measurement), at an angle with respect to the direction of $\gamma_1$, and can be approximated as:

$$R(t) \cong A_{22}^{exp} G_{22}^{exp}(t) \; , \qquad (1)$$

where the factor $A_{22}^{exp}$ is the experimental anisotropy of the $\gamma_1$-$\gamma_2$ cascade and $G_{22}^{exp}(t)$ is the convolution of the theoretical perturbation factor $G_{22}(t)$ with the time-resolution curve of the TDPAC spectrometer. The perturbation factor contains all information about the hyperfine interaction, which in our case is the signature of the EFG tensor, through the measurement of the nuclear quadrupole frequency $\omega_Q$ and the asymmetry parameter $\eta$. For the $^{111}Cd$ probe, with spin $I$ = +5/2 and quadrupole moment $Q$ of the intermediate nuclear state of the $\gamma_1$-$\gamma_2$ cascade , $\omega_Q$ is proportional to the largest component of the diagonalized EFG tensor $V_{33}$ as $\omega_Q = eQV_{33}/40\hbar$. Usually, the experimental results are reported using the quadrupole coupling constant $\nu_Q = eQV_{33}/h$. The diagonalized EFG tensor is completely defined by $V_{33}$ and $\eta = (V_{11} - V_{22})/V_{33}$, using the standard convention $|V_{33}| \geq |V_{22}| \geq |V_{11}|$.

For probes at inequivalent multiple sites $j$ in a polycrystalline sample, sensing *static* electric quadrupole hyperfine interactions (i.e., with constant hyperfine parameters $\omega_Q$ and $\eta$ during the TDPAC measurement), and $I$ = +5/2, the static perturbation factor $G_{22}^{s}(t)$ is:

$$G_{22}^{s}(t) = \sum_j f_j \left( S_{20_j} + \sum_{n=1}^{3} S_{2n_j}(\eta_j) \cos(\omega_{n_j}(\eta_j) t) e^{-\delta_j \omega_{n_j} t} \right) = \sum_j f_j \; G_{22_j}^{s}(t) \; . \qquad (2)$$

The coefficients $S_{2n}$ and the interaction frequencies $\omega_n$ are known functions of $\eta$,[22] being $\omega_n$ proportional to $\omega_Q$ through $\omega_n = g_n(\eta)\, \omega_Q$. The exponential factor accounts for a Lorentzian statistical frequency distribution of relative half-width $\delta$ (in %) around each $\omega_n$, distribution reflecting slight differences in the environment of the probes at site $j$. For single-crystalline samples, the coefficients $S_{2n}$ in Ec. (2) also depend on the orientation of the EFG tensor relative to the detectors' coordinate system. Therefore, the spectrum varies with the sample orientation, thereby allowing determination of the EFG tensor's orientation relative to the crystal axes.[23]

As mentioned earlier, the most commonly used approaches in the literature for addressing the perturbation factor in dynamic HFIs of the ECAE type are those proposed by Bäverstam et al.[17] and Lupascu et al.[11] Both approaches are based on the same premise: the probability that the excited probe atom will decay to a specific final stable electronic state (i.e., unidirectional electronic relaxation) after fluctuations can occur among different charge distributions in the environment of the probe nucleus during its electronic relaxation process.

The BO approach has the advantage of identifying the static or dynamic character for each interaction present in the $R$(t) spectrum, determining the final $EFG_f$, and the relaxation constant $\lambda_r$ (which accounts for the spectrum damping) and the atomic recovery constant $\lambda_g$ (the inverse of the mean lifetime of the electron holes, $\tau_g$, involved in the dynamic process at *each* temperature). On the other hand, the L approach provides information about the initial fluctuating EFGs through a mean $EFG_i$ with an EFG distribution half-width $\delta_i$ that represents the *whole* set of measurements. Instead, the $EFG_f$ and the relaxation rate $\Gamma_r$ to reach this $EFG_f$ are determined for *each* temperature. We note that the parameters $\lambda_g$ and $\Gamma_r$ in the two approaches have the same physical meaning, related to the duration of the electronic recombination. Under certain conditions that will be shown in section II.C, the equivalence between the perturbation factors of the two approaches relates $\lambda_r$ to $\delta_i$, thereby enabling the initial set of fluctuating EFGs ($EFG_i \pm \delta_i$) to be obtained at *each* temperature by applying the BO model to the experiment, as sketched in Figure SI.1 of the Supporting Information. Also, $\lambda_g$ results equal to $\Gamma_r$, and their agreement when analyzing the experiments independently with the two approaches support the proposed equivalence.

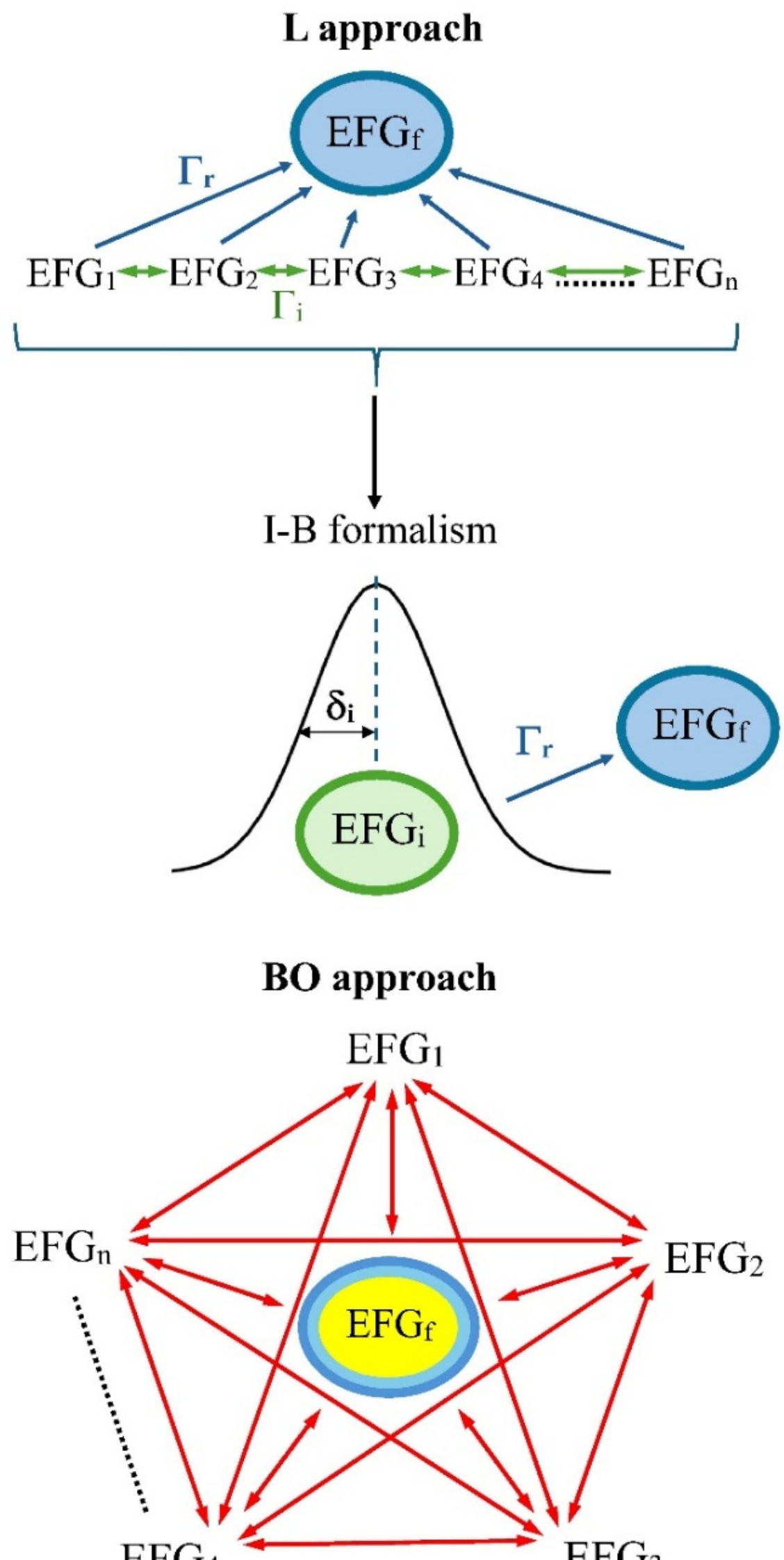

**Figure 1:** Sketch of different scenarios underlying the construction of the perturbation factors $G_{22}(t)$ proposed in the Lupascu *et al.*[11] (L) (upper) and Bäverstam *et al.*[17] (BO) (lower) approaches. The L approach includes a simplification based on the Iwatschenko-Borho *et al.*[18] (I-B) formalism.

### A. The Bäverstam-Othaz (BO) approach

The first approach, proposed by Bäverstam *et al.*,[17] is based on the perturbation factor introduced by Abragam and Pound[24] to analyze rapid, random fluctuations in the probe atom's environment in liquid hosts, which produce rapidly varying EFG tensors. This factor has a purely dynamical nature: $G_{22}^{dyn}(t) = e^{-\lambda_r t}$. This perturbation factor exponentially reduces the spectrum's amplitude toward zero, and the EFG information is completely lost due to rapid random fluctuations among the EFGs detected by the radioactive probe atoms. In the BO approach for solids, the effect of *fluctuating initial* EFGs during the electronic relaxation of the probe atom is represented by the damping introduced by the relaxation constant $\lambda_r$. Additionally, to account for the probability that the excited probe atom decays to a *final stable* electronic state (and thus to a *final stable* EFG, $EFG_f$), BO introduces the atomic recovery constant $\lambda_g$, which is the inverse of the mean *lifetime* $\tau_g = {\lambda_g}^{-1}$ of the trapped electron holes, where the lifetime is measured from the emission of $\gamma_1$, obtaining:

$$G_{22}^{BO}(t) = \sum_j f_j \left[\frac{\lambda_{r_j}}{\lambda_{r_j}+\lambda_{g_j}} e^{-\left(\lambda_{r_j}+\lambda_{g_j}\right)t} + \frac{\lambda_{g_j}}{\lambda_{r_j}+\lambda_{g_j}}\right] G_{22_j}^{S}(t) \quad , \tag{3}$$

where $G_{22_j}^{S}(t)$ is the perturbation factor of eq. 2 describing a *static* HFI at site *j*. The detailed construction of these perturbation factors is outlined in ref. 15. It is important to clarify here that the so-called "electron holes´ (mean) lifetime" is an average over the time interval during which the dynamic regime is detected by each $^{111}$Cd atom contributing to each HFI at a certain temperature. $\tau_g$ actually represents the average lifetime of *only* those electron holes (or fraction of them) that will be finally ionized among all the "dynamic" holes (those involved in the dynamic process) that are trapped at the $^{111}$Cd atom when $\gamma_1$ is emitted (i.e., when the TDPAC measurement begins). In the case of $HFI_u$, *ab initio* calculations will show that the number of "dynamic" and ionized holes is the same; however, this will not be true for $HFI_d$. Performing a least-squares fit of eq. 3 to the experimental $R$(t) spectrum enables the determination of the $EFG_f$ of the final stable state, $\lambda_g$, and $\lambda_r$. It is important to note that the perturbation factor in eq. 3 matches the characteristic shape of the experimental $R$(t) spectra where the ECAE phenomenon is observed. An additional benefit of this method is that $\lambda_g$ and $\lambda_r$ are free parameters in the fitting process for each observed HFI and at *each* temperature, aligning with the fact that both can vary with temperature as they depend on the host's electron availability and/or mobility.[16] It is interesting to observe that, when electronic decay to a final stable state is absent (without turning *off* the dynamic interaction, i.e., $\lambda_g = 0$), the perturbation factor simplifies to:

$$G_{22}^{BO}(t) = \sum_j f_j \, e^{-\lambda_{r_j} t} \, G_{22_j}^{S}(t) \,. \quad (4)$$

Although a final $EFG_f$ does not exist in this potential situation, as might occur in an experimental $R$(t) spectrum, eq. 4 still depends on a well-defined EFG through $G_{22_j}^{S}$, which would be the $EFG_f$ when $\lambda_g$ is small but not null. As proposed in ref. 25, this observation suggests the existence of a favored population in one of the possible charge states, specifically the lowest-energy state. This indicates a preferential population of one EFG among the initial fluctuating EFGs acting on the probe nucleus during the dynamic process, as shown schematically in the lower part of Figure 1. When the switching off of the fluctuating process exists ($\lambda_g \neq 0$), as in most of the ECAE cases reported in the literature, the probe atom ultimately decays to this energetically most stable charge state, resulting in the final $EFG_f$. The physical implications of this fundamental fact ($EFG_i \approx EFG_f$) will be discussed and explained in section II.C after presenting the Lupascu approach.

**B. The Lupascu (L) approach**

On the other hand, the L approach develops a comprehensive numerical simulation of the same physical scenario, based on the Winkler-Gerdau model for dynamic HFIs and stochastic processes.[26, 27] The proposed perturbation factor in these simulations (see eq. 1 of ref. 11) models a unidirectional relaxation toward a single *final stable* $EFG_f$, with a relaxation rate $\Gamma_r$ (probability) that depends on temperature. This occurs after fluctuations among a set of arbitrarily chosen initial EFGs ($EFG_n$) with a unique fluctuating rate $\Gamma_i$. This situation is illustrated in the upper part of Figure 1. These rapid fluctuations, at least visually, erase information about the initial $EFG_n$ from the $R$(t) spectrum, leaving only the final $EFG_f$. When fluctuations among the initial $EFG_n$ are not permitted, the $R$(t) spectrum simulation is affected in the first channels, resulting in additional low-intensity EFGs that conflict with the experimental spectra (see Figures 4 and 7 in ref. 28). In the L approach, comparing simulations with the experimental $R$(t) spectra enables determination of $\Gamma_r$ at each measurement temperature. To obtain $EFG_f$, a static perturbation factor must be fitted to the "static" part of the spectra, i.e., after the spectrum´s strong dampening has finished. It is important to note here that it is $\Gamma_r$ that mainly contributes to the characteristic shape of these simulated spectra: a significant decrease in the signal amplitude in the first nanoseconds and a constant value for longer times (see Figure 2 in ref. 15).

A simplified analytical expression for the perturbation function used in the described simulations can be derived by following the formalism presented by Iwatschenko-Borho *et al.* (I-B),[18] where an *initial* $EFG_i$ relaxes unidirectionally to a *final* $EFG_f$ with the relaxation rate $\Gamma_r$ obtained from previous simulations. Both EFG tensors must be *axially symmetric* (η = 0) and share the *same orientation*. However, to account for the fluctuations between different initial EFGs that create the fluctuating regime before relaxing to the final $EFG_f$, the L approach models this situation by assuming an extensive EFG distribution with a distribution half-width at half-maximum $\delta_i$ centered at a mean *initial* EFG, $EFG_i$ (see middle part of Figure 1), resulting in the L perturbation factor for a single site:

$$G_{22}^{L}(t) = \sum_{n=0}^{3} S_{2n}\,\{[1 - a_n]\cos(n\omega_i t)\exp[-(n\delta_i + \Gamma_r)t] - b_n \sin(n\omega_i t)\exp[-(n\delta_i + \Gamma_r)t] + a_n \cos(n\omega_f t)\exp(-n\delta_f t) + b_n \sin(n\omega_f t)\exp(-n\delta_f t)\} \tag{5}$$

with

$$a_n = \frac{\Gamma_r(\Gamma_r + n\Delta\delta)}{(\Gamma_r + n\Delta\delta)^2 + (n\Delta\omega)^2} \tag{6}$$

and

$$b_n = \frac{\Gamma_r n\Delta\omega}{(\Gamma_r + n\Delta\delta)^2 + (n\Delta\omega)^2} \tag{7}$$

where $\Delta\omega = \omega_f - \omega_i$ and $\Delta\delta = \delta_i - \delta_f$.[11] In these equations, $EFG_i$ and $EFG_f$ are expressed in terms of $\omega_i = 6\,\omega_{Q_i}$ and $\omega_f = 6\,\omega_{Q_f}$.

Equation 5 becomes even simpler when large values of $\Gamma_r$ are considered, as proposed in the L approach (this condition is necessary to observe a static region in the *R*(t) spectra to facilitate the static fit). The *sine* terms can be neglected because, for large $\Gamma_r$, the $b_n$ coefficients approach zero. On the other hand, the factor $\exp[-(n\delta_i + \Gamma_r)t]$ cancels the first term (which contains $\omega_i$), only surviving $G_{22}^{L} = S_{20} + \sum_{n=1}^{3} S_{2n}\, a_n \cos(n\omega_f t)\exp(-n\delta_f t)$ , which is (almost) the perturbation factor for a static hyperfine interaction (it only differs in the $a_n$ coefficient, see eq. 2). At this point, Lupascu *et al.* proposed considering only the dominant term of this summation (which depends on the relative weight of the $S_{2n}$ coefficients). In the case of polycrystalline samples, as in ref. 11, the dominant term is n=1, whereas in single crystals, as in Penner´s experiment,[14] it depends on the sample orientation relative to the experimental setup, with n=2 in this case. The other *n* terms can be ignored because the differences they introduce are small compared to those of the complete static perturbation function. In this way, the coefficient $a_n$ at each temperature can be determined as the ratio of the amplitudes of the "static part" of the *R*(t) spectrum (i.e., for large times) at this temperature relative to that at, generally, high temperature (i.e., when the ECAE remotion is maximum and the spectrum maximizes its amplitude).

Considering all the approximations mentioned in this subsection, it is possible to determine the values of $\Delta\delta$ and $\Delta\omega$ by fitting eq. 6 to the experimentally determined $a_n$ ($a_1$, $a_2$, or $a_3$) at the different measuring temperatures, therefore as a function of $\Gamma_r$ obtained through the numerical simulations at each temperature.[11] In this form, the average mean value of the initial EFGs, $EFG_i$, can be obtained by determining $\omega_i$ ($\omega_i = \omega_f - \Delta\,\omega$), and its EFG distribution determining $\delta_i$ ($\delta_i = \Delta\delta + \delta_f$). Both magnitudes $\omega_i$ and $\delta_i$, related to the initial oscillating EFGs that generate the dynamic interaction, cannot be determined (at least not explicitly) with the BO approach alone.

### C. Equivalence of BO and L perturbation factors

There are some significant differences when applying the L and BO approaches separately to a spectrum. First, the L approach is difficult to use when multiple HFI are present in the spectra, particularly for single-crystalline samples. For example, in Penner´s analysis,[14] this led to the introduction of a large static EFG distribution in the perturbation factor for $HFI_d$, erroneously justified by structural defects, to compensate for part of the strong damping of the spectra, which actually arises from an electronic effect. Second, and critically important, the BO perturbation factor enables detection, in the presence of multiple interactions, of whether each interaction is dynamic or static (as demonstrated in SnO:$^{111}$In).[16] In the L approach, spectra are always fitted with static perturbation factors, even though the hyperfine interactions are dynamic, making it impossible to detect a genuine static component.

It is important to note that although the L and BO approaches are generally regarded as quite different in the literature, it can be shown that under certain conditions, their perturbation factors coincide. In this way, this equivalence will enable a novel, synergistic analysis of the dynamic hyperfine interactions observed in TDPAC experiments, supported by first-principles electronic structure calculations.

To show this, let us consider a single dynamic hyperfine interaction for simplicity. The first approximation is to assume that the mean of the initial fluctuating EFGs, denoted by $\omega_i$ in the L approach, is very similar to the final stable $EFG_f$, associated with $\omega_f$. In this case, $\Delta\omega$ can be neglected and the $b_n$ coefficients vanish (see eq. 7), removing the *sine* terms in eq. 5. The second approximation is to take $n\Delta\delta \rightarrow \Delta\delta$. This assumption is consistent with the fact that, in the static perturbation factor of eq. 2, the exponential factor representing the EFG distribution $\delta$ does not depend on $n$. Hence, both approximations ($n\Delta\delta \rightarrow \Delta\delta$ and $\omega_i \approx \omega_f$) applied to eq. 5 make the coefficient $a_n$ independent of $n$ and allow it to be factored out of the summation. This is a general result valid for all values of $\Gamma_r$. In effect, the condition of large $\Gamma_r$ values mentioned in section II.B was used solely to simplify eq. 5 and facilitate the static fit needed within the L approach. This section aims to show the conditions under which the BO approach and the IB proposal in eq. 5 are equivalent, avoiding any unnecessary approximations used earlier in the L approach.

With these two approximations, eq. 5 is reduced to:

$$G_{22}^{L}(t) = \left\{\frac{\Delta\delta}{\Gamma_r + \Delta\delta}\exp[-(\Delta\delta + \Gamma_r)t] + \frac{\Gamma_r}{\Gamma_r + \Delta\delta}\right\}\sum_{n=0}^{3} S_{2n}\cos(n\omega_f t)\exp(-n\delta_f t)$$

$$= \left\{\frac{\Delta\delta}{\Gamma_r+\Delta\delta}\exp[-(\Delta\delta + \Gamma_r)t] + \frac{\Gamma_r}{\Gamma_r+\Delta\delta}\right\} G_{22}^{S}(t) \ . \qquad (8)$$

Comparing eq. 8 with eq. 3 (just for a single site), we see that both perturbation factors coincide if the relaxation rate $\Gamma_r$ and $\Delta\delta$ ($\Delta\delta = \delta_i$, since $\delta_f$ is really very small) of the L approach are associated with the recovery constant $\lambda_g$ and the relaxation constant $\lambda_r$ of the BO approach, respectively. The equality here between $\lambda_g$ and $\Gamma_r$, which already have the same physical meaning independently in each model, reinforces the assumptions needed to obtain this equivalence.

We note the importance of this equivalence here. By combining the two approaches, it is possible to determine the range of initial EFGs that generate the dynamic interaction, as well as the lifetimes of the electron holes involved in the relaxation process, for *each* measured temperature and *each* observed hyperfine interaction. Within this framework of equivalence, BO directly provides the half-width of the EFG distribution, $\delta_i$, in the dynamic regime at *each* temperature (via $\lambda_r$). Since $\omega_i \approx \omega_f$, and $\omega_f$ is experimentally determined, the BO approach can also provide a reasonably accurate estimate of the mean $EFG_i$ and its distribution half-width, now for *each temperature*. In turn, the L approach allows determination of the central value of the initial fluctuating EFG distribution, $EFG_i$, but it is an average over all measurement temperatures.

After this analysis, it is essential to note that when the results of the L approach in a specific case or experiment confirm the conditions necessary for the equivalence between the two perturbation factors (eqs. 3 and 5), namely $\omega_i \approx \omega_f$, our proposal[16] underlying the BO approach is validated, i.e., that the final $EFG_f$ is among the initial fluctuating EFGs and has the highest probability. In section IV, we will demonstrate that the conditions required for the equivalence are satisfied for the case of $Al_2O_3$:($^{111}$In→)$^{111}$Cd, after our integrated analysis is completed.

## III. *AB INITIO* CALCULATIONS

### A. Calculation details

To obtain a highly accurate description of the electronic density ρ(r), we performed *ab initio* DFT electronic structure calculations for the Cd-doped α-$Al_2O_3$ semiconductor (*corundum* crystal structure). Guaranteeing a diluted-impurity condition is essential for accurately comparing *ab initio* calculations with TDPAC experiments, in which the impurity probe atom is highly diluted (at the ppm level). Details on the crystal structure and the Cd-doped cell used in the calculations are described in section SI.2 of the Supporting Information.

Once an Al atom is replaced by a Cd atom (calling this system $Al_2O_3:Cd^0$), we performed calculations for various charge states of the doped system. Since an electronic recombination process occurs around the $^{111}$Cd impurity in the actual samples, and considering the nominal acceptor nature of $Cd^{2+}$ when it replaces $Al^{3+}$ in the α-$Al_2O_3$ host, we add electronic charge to the $Al_2O_3:Cd^0$ system in 0.05 e− steps up to 1 e−, where $Al_2O_3:Cd^{x-}$ denotes the doped system where an amount of x electrons is added to the system, calling q=x- to the system´s charge state. In addition, we performed calculations removing x electrons to simulate the electron holes created by the Auger process. In these cases, q=x+. The removal and addition of negative charges were compensated by adding a homogeneous background, thereby preserving the cell´s neutrality.[29]

All *ab initio* calculations were performed using the FP-APW+lo method,[30] implemented in the WIEN2k code.[31] Details of the parameters, parametrizations, and the careful treatment and justification for the use of fractional charge states are described in section SI.3 of the Supporting Information.

### B. *Ab initio* results and the origin of dynamic hyperfine interactions

Replacing an Al atom with a Cd one in the $\alpha$-$Al_2O_3$ system ($Al_2O_3$:$Cd^0$) introduces an atomic impurity level into the electronic density of states (DOS) within the energy band gap of pure $\alpha$-$Al_2O_3$ semiconductor, near the top of the valence band (TVB), as shown in Figure 2. This impurity level mainly consists of Cd-4*d*, Cd-5*p,* and O-2*p* states. It is partially filled, with unoccupied states corresponding to one electron, consistent with the nominal acceptor character of $Cd^{2+}$ substituting $Al^{3+}$. In this context, adding electronic charge to the $Al_2O_3$:$Cd^0$ system causes the impurity level's empty states to begin filling—in other words, the Fermi level shifts to the right in a band rigid model—until one electron is added, resulting in $Al_2O_3$:$Cd^{1-}$, which fills the impurity level. Although the calculation adds the charge to the cell of the $Al_2O_3$:$Cd^0$ system, the extra charge mainly localizes at the Cd atom (consistent with the filling of the Cd impurity level in the atom-projected partial DOS[29]) and much less at its ONN, as Figure 3 shows. This electron density $\rho$(r) is a "snapshot" of the added electron [projected on the (010) plane] that fills the impurity acceptor level.

When the charged systems reach their new equilibrium atomic positions, dCd-O1 remains essentially unchanged, while dCd-O2 increases (outward relaxation). In *all* cases, the predicted

EFG has axial symmetry ($\eta = 0$), and the direction of $V_{33}$ stays along the [001] crystalline axis. This information about the EFG tensors for different charge states, previously unknown, will be crucial for validating the requirement, within the L approach, that the initial and final EFGs must have *the same orientation and axial symmetry*, as needed by the I-B formalism.[18]

Figure 4 compares the *ab initio* predictions of $V_{33}$ with the experimental results obtained by Penner *et al*.[14] The dashed areas indicate the range of $V_{33}$ values for each interaction across the entire measured temperature range.

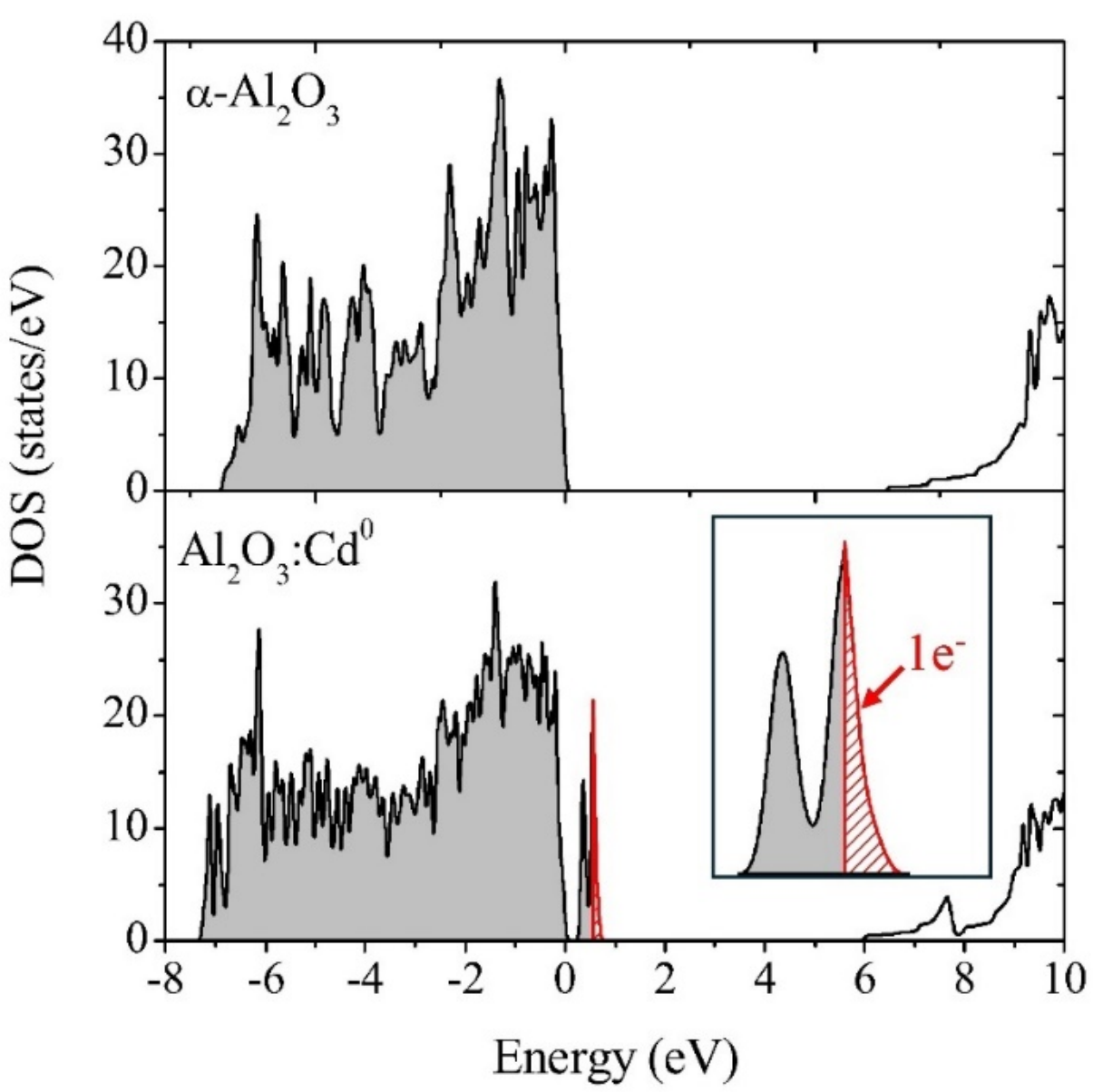

**Figure 2:** Total electronic density of states (DOS) for pure (top) and Cd-doped (bottom) α-$Al_2O_3$ ($Al_2O_3$:$Cd^0$). The gray areas indicate occupied states. A zoom-in on the Cd impurity level is included. The dashed red region highlights the acceptor level introduced by the Cd impurity.

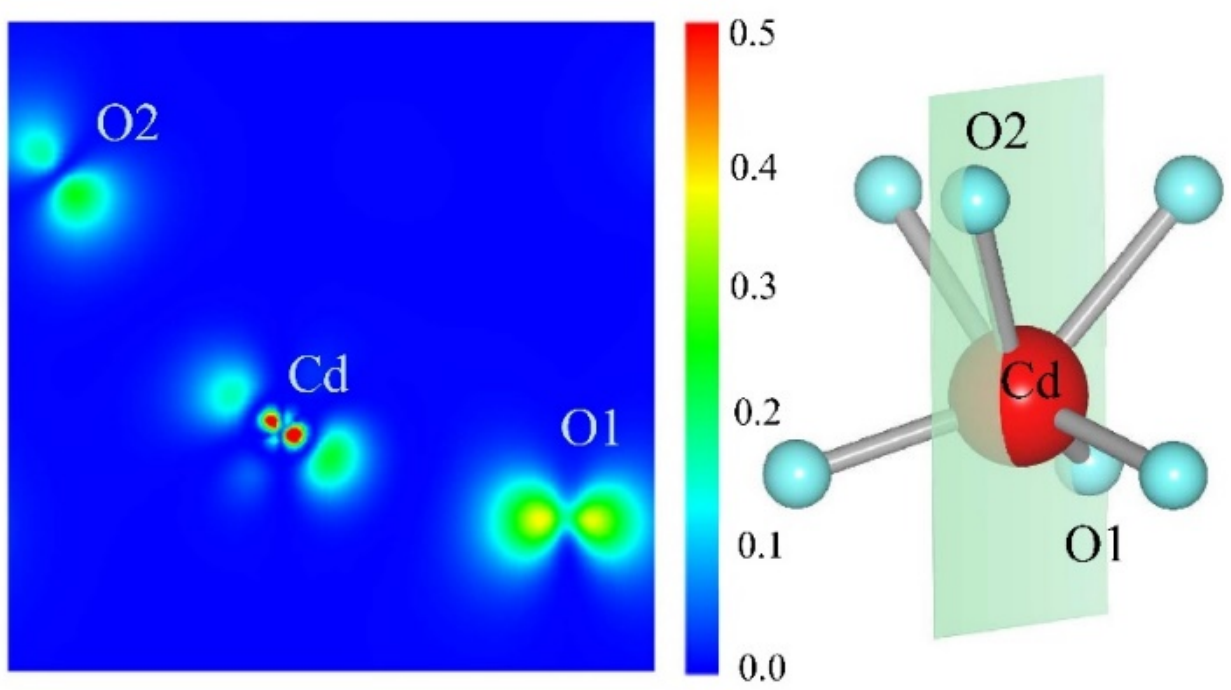

**Figure 3:** Electron density $\rho(r)$ in the $Al_2O_3:Cd^{1-}$ system, corresponding to the energy range of the added electron in the DOS, projected onto the (010) plane. The right side shows the Cd-O1 and Cd-O2 bonds in this plane.

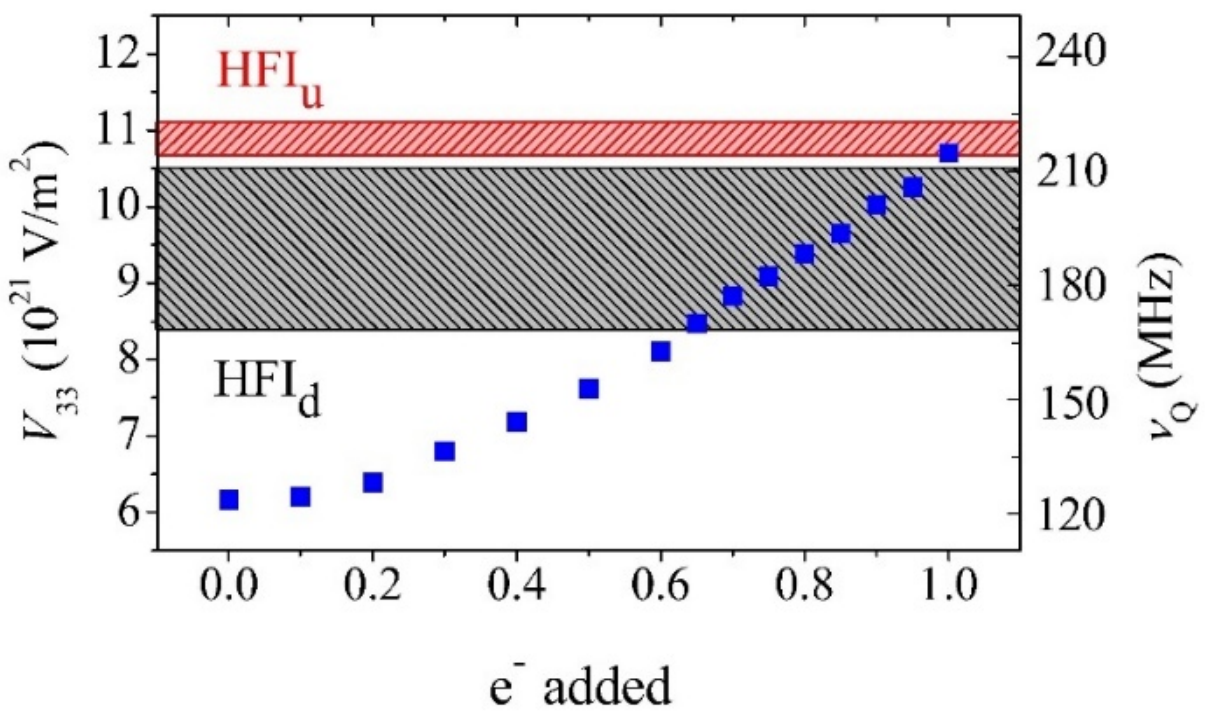

**Figure 4:** Predicted $V_{33}$ at the Cd site in $Al_2O_3:Cd^{x-}$ as a function of added electrons to the system (blue squares), using lattice parameters at room temperature. The dashed areas indicate the experimental values of $HFI_u$ (red) and $HFI_d$ (gray) in the 4-973 K measurement temperature range. To obtain $V_{33}$ from $\upsilon_Q$, we used $Q = +0.83(13)$ b.[32]

Figure 4 shows that $HFI_u$ can be associated with $^{111}Cd$ probes located at defect-free substitutional Al sites, with the impurity level fully occupied ($Al_2O_3:Cd^{1-}$ system), indicating the absence of trapped electron holes. The duplication of $V_{33}$ when one electron is added to the $Al_2O_3:Cd^0$ system, which ionizes the electron hole mainly trapped at the Cd atom, can be qualitatively explained by examining the electronic density $\rho(r)$ shown in Figure 3. This change in the EFG results from the modification of $\rho(r)$ inside the muffin-tin sphere around Cd, especially near the Cd nucleus. To analyze this change, we consider how the magnitude and sign of $V_{33}$ depend on the angle between a negative charge distribution and the $V_{33}$ direction (as described by eq. 14 in ref. 15). Negative charges at angles less than 54.7° from the $V_{33}$ direction (parallel to the [001] crystal axis), shown as the red spot on the left side of the Cd atom in Figure 3, contribute negatively to $V_{33}$. Those closer to the (001) plane (at an angle less than 35.3° from this plane) contribute positively, as does the larger red spot to the right side of Cd. The greater accumulation of negative charge at the Cd site closer to the (001) plane accounts for the notable increase in EFG as negative charge is added to the system.

Penner´s results show that the reported $V_{33}$ value for $HFI_u$ is not constant, but decreases slightly as T increases.[14] To understand this temperature dependence, we performed calculations in the $Al_2O_3:Cd^{1-}$ system, using experimental lattice parameters that change with temperature.[33] The strong agreement between the experimental data and our calculations of the $V_{33}$ temperature dependence (see Figure 5) indicates that this behavior arises from thermal lattice expansion rather than electronic effects. Notably, the significant nonlinear expansion of the *a* and *c* lattice parameters with increasing temperature,[33] as used in the calculations, explains the slight linear dependence of $V_{33}$ observed.

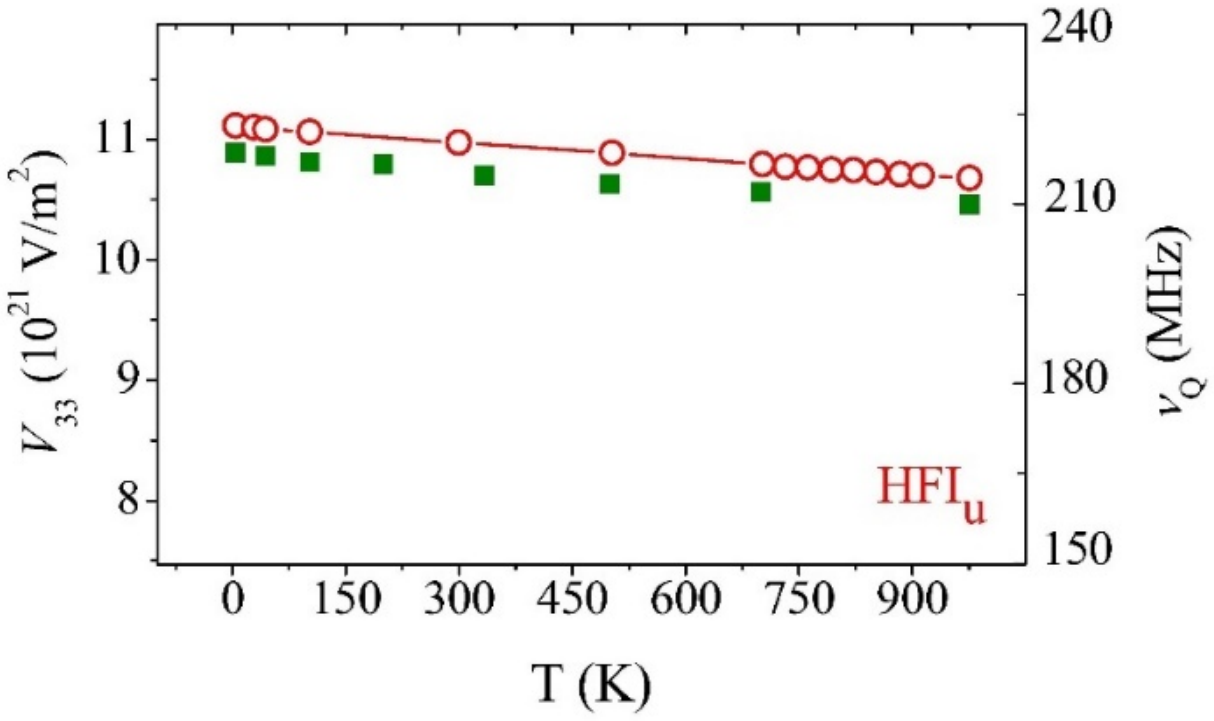

**Figure 5:** Comparison of predicted $V_{33}$ for $Al_2O_3$:$Cd^{1-}$ (green squares) using the experimental lattice parameter as a function of temperature with the experimental $V_{33}$ values of HFI$_u$ (open red circles) reported by Penner *et al.*[14]

Conversely, as demonstrated by the *ab initio* EFG calculations, HFI$_d$ can be connected to $^{111}$Cd probes at substitutional Al sites. Still, the final stable electronic environments vary with temperature. In fact, both Figures 4 and 6 show that the $V_{33}$ values of HFI$_d$ as a function of temperature match very well with the $V_{33}$ predictions within the range from q=0.65- to q=1-. Figure 6 also demonstrates that the complete electronic recovery of the $^{111}$Cd atom (filled impurity level) occurs at very high and very low temperatures, where HFI$_d$ converges, not by chance, to the value of HFI$_u$.

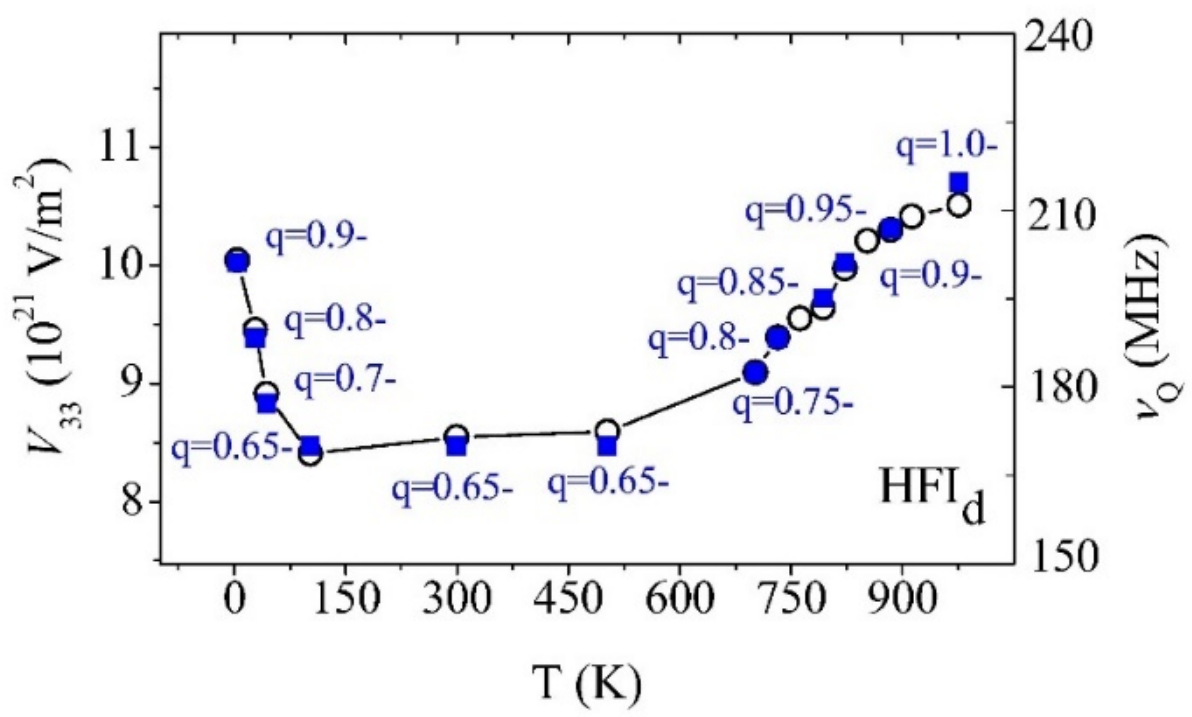

**Figure 6:** Comparison of predicted $V_{33}$, for different q values using lattice parameters at RT (blue squares), with the experimental $V_{33}$ values of $HFI_d$ (open circles) reported by Penner *et al.*[14] as a function of temperature.

But why might these specific EFGs of $HFI_d$ have been observed in the experiment? Let's analyze the defect-formation energy study of a Cd impurity in α-$Al_2O_3$ as a function of the charge state (see section SI.4 of the Supporting Information for details).

Figure 7 shows the results for the defect formation energy across all studied charge states, between q=0 and q=1- (red lines). It also displays the result of adding 1.1 e-, which is 0.1 e- more than in the system with the impurity level filled (dash blue line), and with two electron holes (q=1+), meaning one electron removed from the $Al_2O_3$:$Cd^0$ system (green line). These results indicate that adding more electrons when the impurity level is filled is energetically less favorable, as well as the existence of two electron holes trapped at the Cd atom (unless the Fermi energy $\varepsilon_F$ is very close to the top of the VB or in deeper states).

As shown, in an acceptor case such as the current impurity-host system, the charge state corresponding to the fully occupied impurity level (q=1-) has the lowest energy over a wide range of the Fermi energy (see Figure 7a). However, when an EC creates the probe atom, subsequent

Auger processes generate many electron holes that diffuse toward the atom´s outer shells (within about $10^{-14}$ s), leaving it highly ionized. These holes are then filled based on the defect formation energy. Because the initial stage of the electronic recovery of the Cd atom – where electrons from neighboring atoms fill the holes - is very fast (about $10^{-12}$ s), only a few electron holes in the outermost shells can eventually be detected in a TDPAC experiment. This may produce either a dynamic or a static HFI, depending on the electrical properties (metallic, semiconducting, or insulating) of the host and the lifetime of the nuclear level that emits $\gamma_1$.[16]

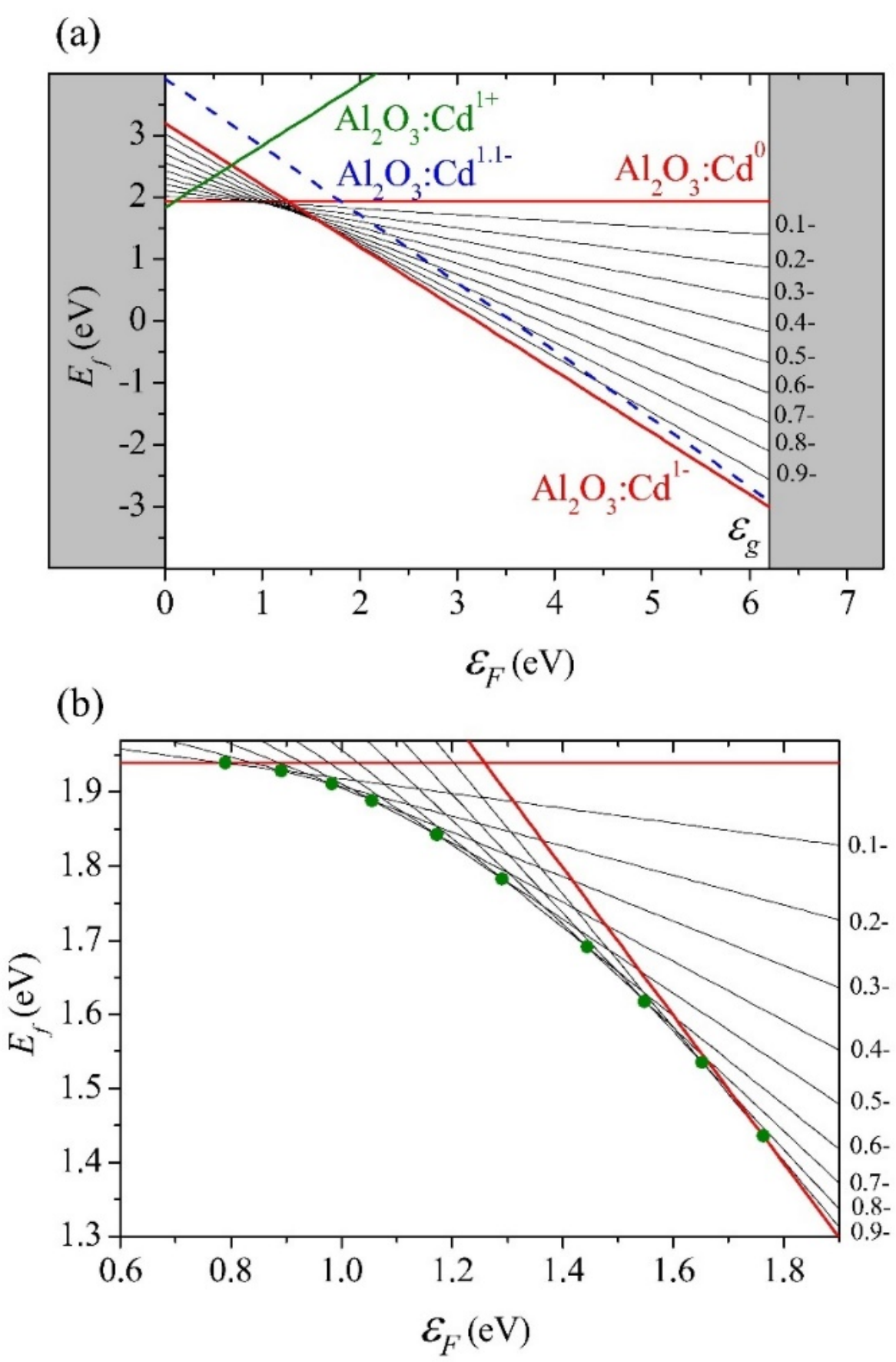

**Figure 7:** (a) Defect formation energy $E_f$ as a function of Fermi energy $\varepsilon_F$ for all studied $Al_2O_3:Cd^q$ systems. The gray energy regions below 0 eV and above $\varepsilon_g$ represent the valence and conduction

bands, respectively, as predicted for pure α-$Al_2O_3$. (b) Zoomed-in view of the band gap energy region, where lines for q=n and q=n+0.1 intersect.

Taking all this into account, it is now clear where $HFI_u$ originates. Its EFG value matches the EFG prediction for $Al_2O_3$:$Cd^{1-}$, indicating that it corresponds to $^{111}$Cd atoms located at structurally defect-free cation sites. These atoms, after experiencing fluctuating electronic environments caused by shifts during the filling of the last empty Cd electronic states, decay to the stable charge state corresponding to the impurity level being fully filled. Inspecting Figure 7a and then Figure 7b for greater accuracy, we observe that systems with charge states between q=0 and q=1- (with different predicted EFGs) have very similar formation energies when the Fermi energy is around $\varepsilon_F$ = 1.25 eV; the same applies for q=0 and q=1+ near the TVB, supporting the idea of random fluctuations in this charge state range and, consequently, between their EFGs. These fluctuations are necessary in all approaches proposed to date for describing the ECAE phenomenon and constructing their perturbation factors, as previously described.

Now, let's analyze the origin of $HFI_d$. The previous situation can be slightly altered if the "reservoir of electrons" in the current sample decreases the Fermi energy to below approximately 1.8 eV. This variation is generally related to the balance between intrinsic donor and acceptor impurities, the change in electron availability due to thermal effects, and, as seen in this case, electron conductivity at very low temperatures. The ECAE process favors this decrease and is temperature-dependent. Figure 7b shows that as the Fermi energy drops, charge states with a lower proportion of added electrons have the lowest energy. The final EFGs observed by $HFI_d$ as a function of temperature match perfectly with $^{111}$Cd atoms decaying to a final stable electronic configuration, corresponding to the addition of 0.65 to 1 electron to the system, which changes the occupation of the impurity level from 65 to 100% (see Figure 4). This corresponds to a Fermi energy range of approximately 1.5 to 1.8 eV across the entire measurement temperature range (see Figure 7b). In Figure 6, the temperature dependence of the Fermi energy can be deduced, thereby confirming the system's final stable charge states. In effect, the lower Fermi energy value ($\varepsilon_F$ = 1.5 eV and q = 0.65-) aligns with the very high population observed for $HFI_d$ in ref. 14, between approximately 75 and 500 K, indicating that there are not enough electrons to fully fill the impurity level for almost all the $^{111}$Cd probes due to the additional electron holes created by the EC and the low electron availability at these temperatures. In TDPAC experiments on $\alpha$-$Al_2O_3$ single crystals implanted with $^{111m}Cd^+$ ions, a *single static* HFI was observed (between 77 and 873 K) with the same EFG of $HFI_u$.[13] In these experiments,[13] the same sensitive nuclear state of $^{111}$Cd is used to measure the HFI. However, this state is populated from the $11/2^+$ isomeric state (with a very long lifetime of 48.6 minutes) of $^{111m}$Cd via a 151 keV $\gamma$-decay, without producing Auger cascades. This experimental result demonstrates that, in this oxide, in the absence of Auger electron holes trapped at Cd beyond the single hole at the acceptor impurity level, all $^{111}$Cd probes fill the impurity level (q=1-) at all temperatures, aside from the time required to reach this stable state. This supports the conclusion that the EC is a necessary condition for observing the final stable $HFI_d$ values in this oxide. However, relying on the absence of "aftereffects" when using $^{111m}$Cd to assess whether the EC is a necessary condition for observing dynamic HFIs is incorrect. In this case, there are no dynamic HFIs because of the long lifetime of the 11/2+ state, not because of the absence of the EC. As stated in ref. 16, without EC, any probe impurity that introduces acceptor or donor states (in semiconductors and insulators) and whose ionization changes the EFG should exhibit a dynamic HFI, as long as the lifetime of the dynamic process (such as the ionization of electron holes or donor electrons, respectively) exceeds the lifetime of the nuclear levels that feed the $\gamma_1$ state and the state itself.

The current electronic interpretation of $HFI_d$ differs entirely from the structural defect scenario proposed by Penner *et al.*, as discussed in section II.C. Is it possible to have more than one electron hole as a final state for $HFI_d$? For this to occur, the Fermi energy must be very low, and these charge states should have a lower formation energy than the $Al_2O_3$:$Cd^0$ state. To verify this, we calculated the formation energy for two holes ($Al_2O_3$:$Cd^{1+}$). As shown in Figure 7a, this situation is indeed energetically unfavorable. The same trend, gradually, applies to any system between $Al_2O_3$:$Cd^0$ and $Al_2O_3$:$Cd^{1+}$.

## IV. BÄVERSTAM-OTHAZ ANALYSIS OF THE EXPERIMENT AND INTEGRAL DISCUSSION

We present the results of applying the BO approach to the TDPAC experimental data reported by Penner *et al.*[14] Figure 8 shows how well eq. 3 fits the selected *R*(t) spectra measured at 4 K, 293 K, and 910 K. As expected, two hyperfine interactions are required to account for the spectra at each temperature, with coupling constants and an asymmetry parameter of zero identical to those reported for $HFI_u$ and $HFI_d$ in ref. 14. In our case, both interactions were treated as single-crystalline. The $V_{33}$ orientation for both interactions was determined to be parallel to the [001] crystal axis, as reported by Penner *et al.* for $HFI_u$. The fitted populations are similar to those in Penner's analysis. However, the frequency distributions, especially for $HFI_d$, are notably lower because a dynamic perturbation factor is used instead of the static perturbation factors in ref. 14, as required by the L approach. The BO approach provides an excellent fit across the entire spectrum, particularly within the first 5-15 ns. In contrast, a static perturbation factor cannot accurately fit these short times.

As discussed in section II, applying the BO approach to a single experimental spectrum enables determination of the final EFG for $HFI_u$ and $HFI_d$, as well as the atomic recovery constant $\lambda_g$ and the relaxation constant $\lambda_r$, for each interaction at a given temperature.

Figure 9 compares the relaxation rate $\Gamma_r$ for $HFI_u$ as a function of temperature, derived from Penner *et al.*'s data using the L approach, with $\lambda_g$ (shown in red) obtained here via the BO approach from the $R$(t) spectra in Figure 8. As shown, the results from both approaches agree perfectly at these three representative temperatures. Having identified $\Gamma_r$ with $\lambda_g$ (the inverse of the electron hole lifetime $\tau_g$), its temperature dependence indicates an increase in electron availability at higher temperatures. At very low temperatures, this is likely due to increased electron conductivity. The different slopes in the $\Gamma_r$ dependence on T (below 100 K and above 750 K) are examined in detail through two transport models for electrons in ref. 14. The excellent agreement shown in Figure 9 provides the first experimental evidence of the equivalence between the BO and L approaches, as $\Gamma_r$ and $\lambda_g$ represent the same physical quantity but are obtained through very different methods.

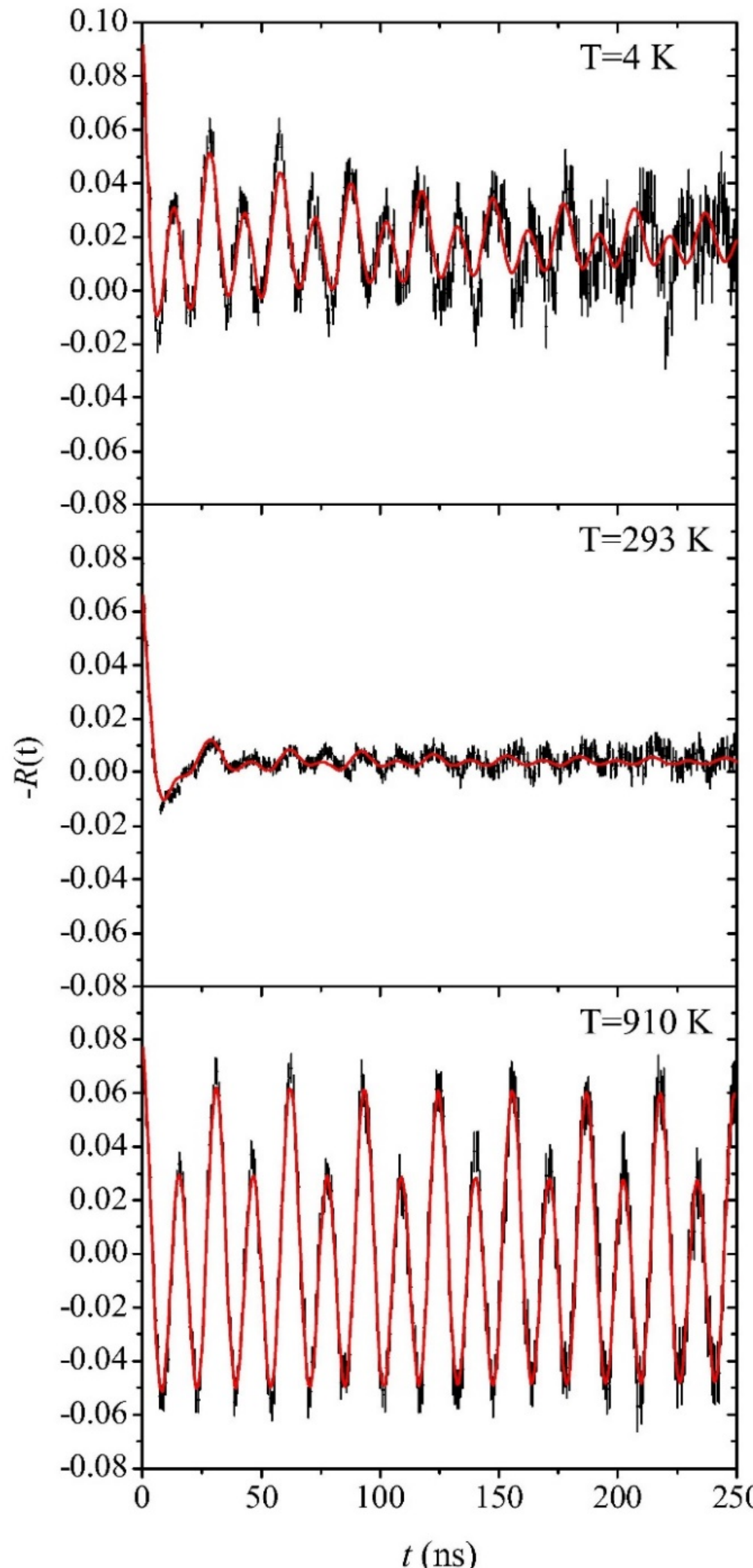

**Figure 8:** Fits to the *R*(t) spectra measured at 4 K, 293 K, and 910 K by Penner *et al.*[14] using the BO perturbation factor implemented in the *TDPAC* fitting program,[34] modified for single crystals.

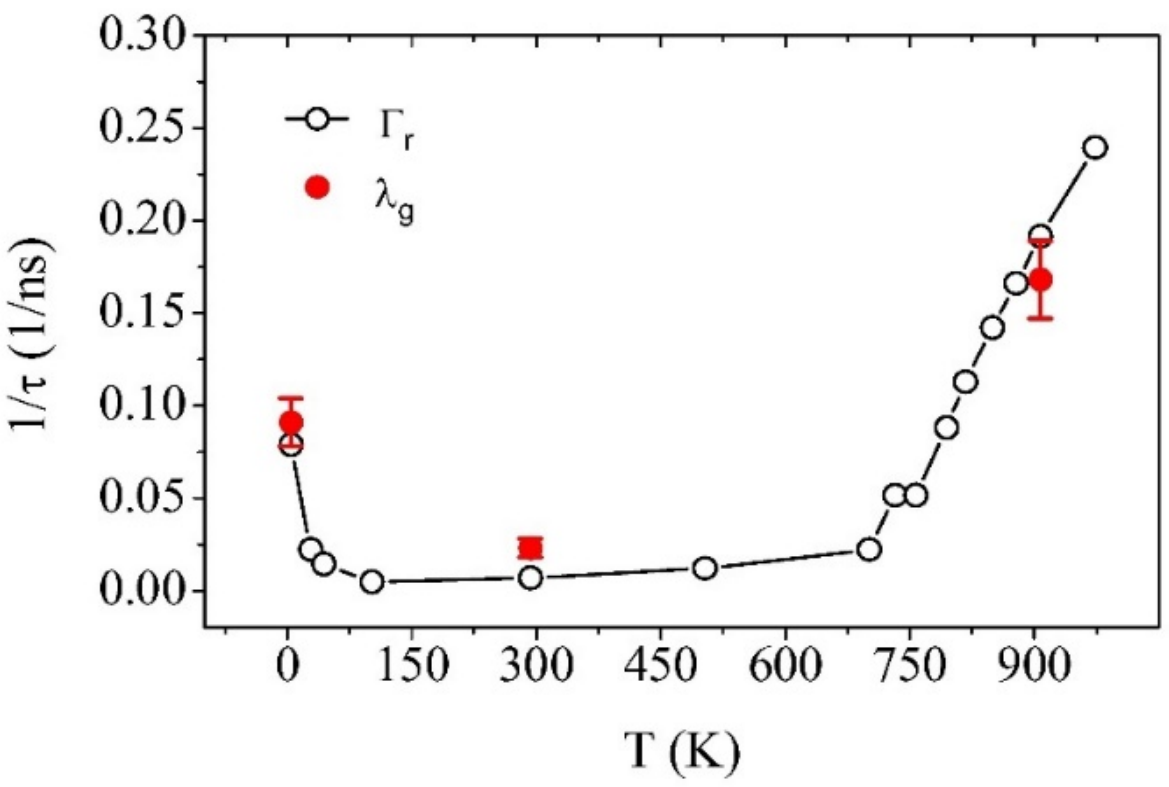

**Figure 9:** Comparison of the relaxation rate $\Gamma_r$ (open circles) as a function of measurement temperature, obtained by Penner *et al.* applying the L approach to $HFI_u$,[14] and the atomic recovery constant $\lambda_g$ for this interaction determined here with the BO approach (red filled circles).

In Table I, we present the dynamic parameters $\lambda_g$ and $\lambda_r$ for both $HFI_u$ and $HFI_d$, along with their populations and the electron hole lifetime $\tau_g$ for each interaction. These values were determined by fitting the BO perturbation factor to the *R*(t) spectra shown in Figure 8.

**Table 1:** Dynamic hyperfine parameters and the fraction *f* of $^{111}$Cd probes detecting each interaction, $HFI_u$ and $HFI_d$, determined by applying the BO approach to the three selected *R*(t) spectra shown in Figure 8.

| | $HFI_u$ | | | | $HFI_d$ | | | |
|---|---|---|---|---|---|---|---|---|
| T (K) | *f* (%) | $\lambda_g$ (MHz) | $\lambda_r$ (MHz) | $\tau_g$ (ns) | *f* (%) | $\lambda_g$ (MHz) | $\lambda_r$ (MHz) | $\tau_g$ (ns) |
| 4 | 48 (1) | 91 (13) | 2.2 (4) | 11 (2) | 52 (1) | 22 (6) | 39 (5) | 45 (12) |
| 293 | 18 (1) | 23 (5) | 109 (8) | 43 (9) | 82 (1) | 48 (14) | 202 (25) | 20 (6) |
| 910 | 90 (1) | 168 (21) | 4.7 (7) | 6 (1) | 10 (1) | 192 (23) | 8 (2) | 5 (1) |

As mentioned in section II.C, the $\lambda_r$ parameter is associated with $\Delta\delta$ $(= \delta_i)$ of the L approach. At this point, we must note that the value $\Delta\delta^u$ = 90 MHz (of $HFI_u$) reported in ref. 14 was determined for n = 1 using eq. 6, even though the single-crystal orientation in the experimental setup requires n = 2. Therefore, $\Delta\delta^u$ = 45 MHz (HWHM) and $\Delta\upsilon_Q^u$ = 10 MHz are the correct values that fit eq. 6 perfectly, with n = 2, to the experimental data shown in Figure 3 of ref. 14. In the L approach, a complete set of measurements as a function of temperature is required to determine these "average" values for $\delta_i^u$ and $\nu_{Qi}^u$, representing the entire temperature range. Hence, $\delta_i^u$ = 45 MHz represents a wide temperature range of initial EFGs´ distributions. In comparison, the BO approach yields $\lambda_r$ values across temperatures ranging from 2.2 to 109 MHz (average of 55.6), further demonstrating the agreement between the two methods.

By leveraging the BO approach´s ability to determine all dynamic parameters listed in Table I for both interactions and temperatures (including the hyperfine parameters not shown), its equivalence to the L approach, and by combining these results with the *ab initio* predictions of the EFG as a function of the impurity´s charge state shown in Figures 4 and 6, we can describe the electronic behavior of the dilute system during the dynamic regime across the entire temperature range.

We found that the maximum number of dynamic electron holes (those present when $\gamma_1$ is emitted) in $HFI_u$ is lower than in $HFI_d$ at all temperatures. At 293 K, this number reaches up to one hole in $HFI_u$ and at least five holes in $HFI_d$. The maximum number of electron holes was determined by correlating the ranges of the initial EFGs with the corresponding charge states q (see Figure 10). In turn, the initial EFGs´ range at each temperature was calculated as $EFG_i \pm \delta_i$, considering $\delta_i = \lambda_r$ and $EFG_i \approx EFG_f$ within the equivalence.

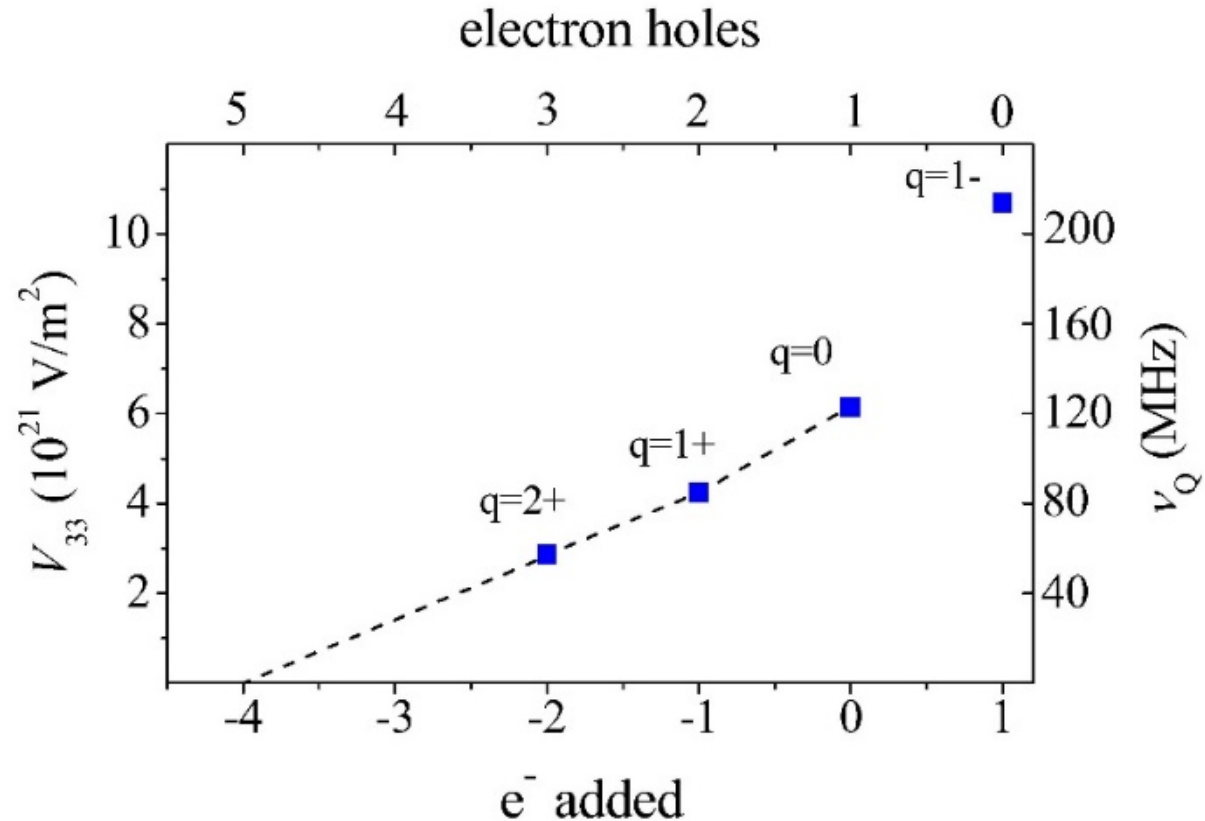

**Figure 10:** Predicted $V_{33}$ for $Al_2O_3$:$Cd^q$ without adding electrons (q=0), adding one electron (q=1-), and removing one (q=1+) and two (q=2+) electrons. The upper horizontal axis indicates the number of simulated electron holes.

As shown in Table I, at 293 K (where the *R*(t) spectra are strongly damped, and the ECAE are stronger), electronic availability is very low. At this temperature, before the $\gamma_1$ emission, the deepest electron holes at the probe atoms are ionized in two distinct ways. When $\gamma_1$ is emitted, 18 % of probe atoms ($HFI_u$) have one trapped electron hole, while 82 % of the probes ($HFI_d$) retain at least five electron holes. This is the primary physical difference between the two HFIs, not their final electronic state, which depends on the availability of electrons at each temperature. Now, let us describe the behavior of electron holes in the dynamic regime at this temperature. As shown in Table I, $HFI_u$ takes longer than $HFI_d$ to fill fewer electron holes. Remember that $\tau_g$ starts when $\gamma_1$ is emitted and ends when the dynamic regime terminates. Since deeper empty states must be filled first to reduce the system's energy, $HFI_d$ uses the first, let's say, 15 ns to fill four electron holes, then five ns to reach the final q= 0.65+ stable charge state, totaling 20 ns. On the other hand, during these first 15 ns, $HFI_u$ does not fill any electron holes because its four deeper electron holes are already filled before the $\gamma_1$ emission. When $HFI_d$ begins to fill the last holes, $HFI_u$ starts filling its remaining electron hole, at the same rate as $HFI_d$ in the first five ns, and fills it in the following 23 ns, reaching the q=1- final stable charge state.

At 4 K, electron availability decreases further. Despite this, $\lambda_r$ drops significantly - by a factor of 50 for $HFI_u$ and 7 for $HFI_d$ - restoring the spectral amplitude, an effect previously observed in TDPAC experiments on other oxides doped with $^{111}$In ions.[11, 12, 35-37] At this temperature, the distribution of initial oscillating EFGs for $HFI_d$ is roughly 20 times wider than for $HFI_u$ (see $\lambda_r$ in Table I). In this very low-temperature range, Penner *et al.* attribute the higher relaxation rates $\Gamma_r$, which indicate shorter hole lifetimes, to increased electron mobility resulting from carrier transport across localized states of the same energy, known as coherent tunneling.[38] In 11 ns, about half of the probes sensing $HFI_u$ completely fill the impurity level with a small quantity of negative charge, whereas those sensing $HFI_d$ require three times longer to reach the q = 0.9- charge state. Considering the dynamic hyperfine parameters measured for $HFI_u$, which produce an almost undamped contribution and are sensed by only half the probes, the damping observed in the $R$(t) spectrum at this temperature is due to $HFI_d$.

At high temperatures (T = 910 K), the two interactions are similar because $HFI_d$ tends toward the most populated $HFI_u$. Increased electron availability due to thermal effects (a high Fermi energy) ensures that 90% of $^{111}$Cd probes fill the impurity level during the dynamic process. For the same reason, both interactions have very low $\lambda_r$ values because of a narrow range of initial EFGs (i.e., low variation in initial electronic configurations) and very small fractions of initial electron holes during the dynamic regime (very close to the q = 1- charge state). Accordingly, at this temperature, the hole lifetimes $\tau_g$ for both interactions are also very short, with 10% of the probes ($HFI_d$) filling quickly ($\tau_g^d$ = 5 ns) the small fraction of dynamic electron holes up to a charge state q = 0.95-. One nanosecond later ($\tau_g^u$ = 6 ns), 90% of the probes ($HFI_u$) complete filling of the impurity level, thanks to the favorable energetic situation for the q=1- charge state (see Figure 7a) due to the high Fermi energy at this temperature and the lower fraction of initial electron holes trapped at the $^{111}$Cd probes sensing $HFI_u$.

From this electronic behavior, we can conclude that as the number of electron holes during the dynamic regime (and thus the range of oscillating initial EFGs) increases, a larger fraction of them contributes to producing the final $EFG_f$. This applies to $HFI_d$ transitioning from high to intermediate temperatures. Conversely, when the initial number of electron holes is small (e.g., at high temperature for $HFI_d$), the range of oscillating initial EFGs becomes limited, and the final $EFG_f$ tends to match that measured for $HFI_u$ (with the impurity level filled).

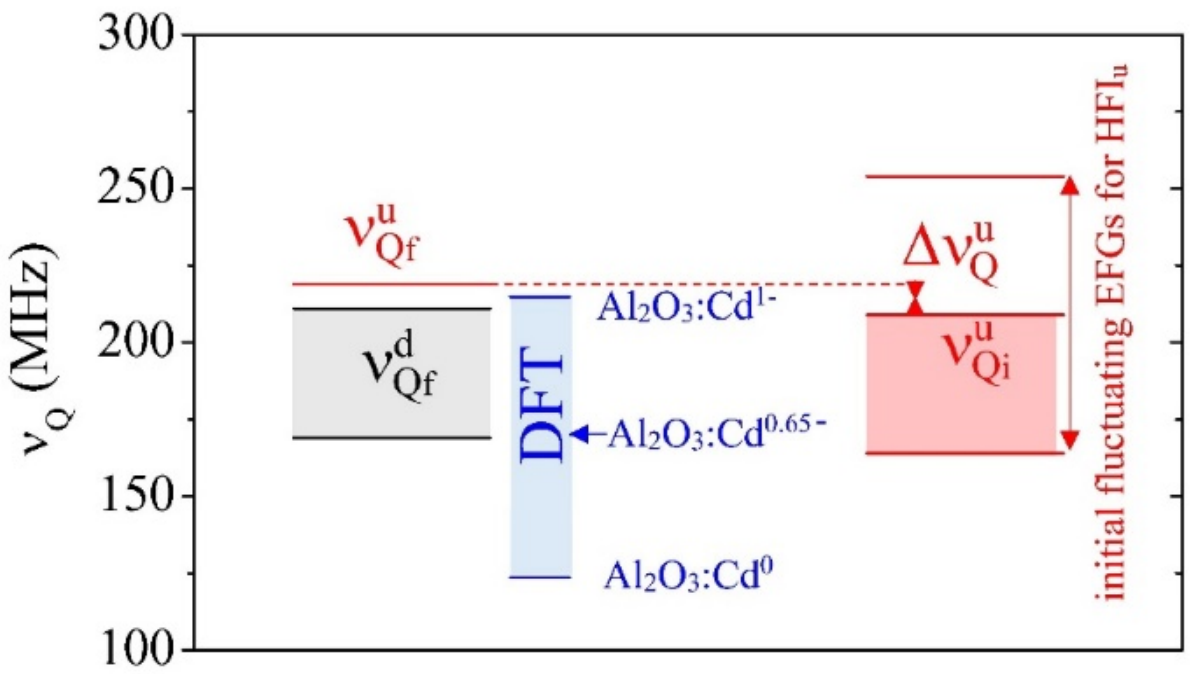

**Figure 11:** Comparison between the final stable coupling constant $\upsilon_{Q_f}$ reported by Penner *et al.*[14] in the 4-973 K temperature range for both hyperfine interactions $HFI_u$ and $HFI_d$ (on the left), along with *ab initio* predictions for $\nu_Q$ as a function of the charge states of the Cd atom (middle). On the right, the average value (across all measured temperatures) of the initial fluctuating coupling constant, $\nu^{u}_{Q_i}$, and its distribution width (L approach) are shown for $HFI_u$, together with $\Delta\nu^{u}_{Q} = \nu^{u}_{Qf}$- $\nu^{u}_{Qi}$ .

Let´s now present a graphical summary and discuss the proposed scenario for the origin of the EFG in more detail, including the initial stages during the dynamic regime and the final stable state after fluctuations cease, for $HFI_u$ and $HFI_d$ within the framework of the BO and L approaches, as well as *ab initio* predictions. Figure 11 shows, on the left, the experimental EFGs observed for $HFI_u$ and $HFI_d$ across the entire temperature range (expressed for comparison purposes in terms of the coupling constant $\nu_Q$, see section II); in the middle, the *ab initio* predictions of the EFG at the Cd site for charge states from q=0 to q=1-; and on the right, the initial mean value $\nu_{Qi}^{u}$ and its distribution half-width $\delta_i^{u}$, as determined by Penner *et al.* using the L approach. As described in section III.B, the predicted EFG for q=1- perfectly matches the experimental EFG of $HFI_u$ (its final stable EFG), and the range of the experimental EFGs (the final stable ones) for $HFI_d$ over the entire temperature range corresponds to predictions between q=0.65- and q=1-. Predicted EFGs that simulate more than one electron hole do not match the experimental values reported in ref. 14, falling significantly below this range. For example, with two electron holes (q=1+), the predicted value is $V_{33}$=4.23 x$10^{21}$ V/m$^2$ (84.9 MHz). Refer to Figures 4, 6, and 10 for comparison. Additionally, this situation is energetically unfavorable (see the green line in Figure 7a). The same scenario occurs when more than one electron is added to the system, predicting $V_{33}$ = 11.10 x $10^{21}$ V/m$^2$ and $V_{33}$ = 11.43 x $10^{21}$ V/m$^2$ for q = 1.1- and q = 1.2-, respectively, which exceed the $HFI_u$ experimental value (see Figures 4 and 5). Similarly, for these charge states, the defect formation energy begins to increase (dashed blue line in Figure 7a), leading once again to an unfavorable outcome.

In Figure 11 (on the right), due to the uncertainty of the EFG sign in a standard TDPAC experiment, we assume the positive sign of $\nu_{Q_f}^{u}$ given by the *ab initio* calculation. Since eq. 6 is insensitive to the $\Delta\nu_{\mathrm{Q}}^{\mathrm{u}}$ sign, we select the smaller value for $\nu_{\mathrm{Q}i}^{\mathrm{u}}$ between the two options from the determination of $\Delta\nu_{\mathrm{Q}}^{\mathrm{u}} = \nu_{\mathrm{Q}f}^{\mathrm{u}} - \nu_{\mathrm{Q}i}^{\mathrm{u}}$ , because the other would lead to an initial positive value $\nu_{\mathrm{Q}i}^{\mathrm{u}}$ larger than $\nu_{\mathrm{Q}f}^{\mathrm{u}}$, which in the calculations corresponds to the addition of more than one electron. This is not only energetically unfavorable but would also yield unphysical initial EFGs, arising from adding electrons to the conduction band rather than from electron holes created by the EC. For the same reason, we have only highlighted the initial EFGs region in this figure corresponding to $\nu_{\mathrm{Qi}}^{u} - \delta_{i}^{\mathrm{u}}$, since the majority of the other side of the frequency distribution (above $\nu_{\mathrm{Q}f}^{\mathrm{u}}$) appears unphysical in this case. As shown, the range of initial EFGs that originate the dynamic regime between $\nu_{\mathrm{Q}i}^{\mathrm{u}}$ and $\nu_{\mathrm{Qi}}^{u} - \delta_{i}^{\mathrm{u}}$ (shaded pink region on the right side of Figure 11) using the L approach matches perfectly with the range of experimental final EFGs for $\mathrm{HFI_d}$ across the entire measurement temperature range (shaded grey region on the left side of Figure 11). This agreement strongly supports the interpretation of $\mathrm{HFI_d}$'s final EFGs, which are accurately reflected by the charge states between q=1- and q=0.65-. On the other hand, the range of initial EFGs for $\mathrm{HFI_u}$, which we note is an average value across all measurement temperatures obtained by Penner *et al.* within the L approach (with a distribution half-width of about 45 MHz), is roughly the same magnitude as that determined in the BO approach (using $\lambda_r$) at intermediate temperatures for this interaction (see Table I), where the dynamic effects are more pronounced. Assuming the equivalence between the two approaches is valid, which requires that $\mathrm{EFG_i} \approx \mathrm{EFG_f}$ (as shown in Figure 11 on its right side), the range of the initial EFGs for $\mathrm{HFI_u}$ can be determined from the experimental final $\mathrm{EFG_f}$ to lower values. Thus, the range of initial EFGs determined by the BO approach at RT (with a spread of $\lambda_r$ = 109 MHz) corresponds to the range of predicted EFGs between the situation in which the impurity level is filled and that of having one electron hole (see Figure 11, middle, shaded light blue region). A similar situation was observed with

SnO:$^{111}$In($\rightarrow$$^{111}$Cd), where the dynamic regime resulted from charge-state fluctuations involving up to one electron hole [16]. In our system, the $\lambda_r$ values for $HFI_u$ at high and very low temperatures are minimal, indicating that the possible electronic configurations during the dynamic regime correspond to a narrow charge-state range very close to the q = 1- state.

Unfortunately, in the case of $HFI_d$, it was not analyzed with a dynamic character by Penner *et al.*[14] Instead, it was artificially treated as a *polycrystalline static* interaction. However, the BO approach and the integral analysis presented here enable the determination of the initial EFGs range at each temperature for this interaction. For this, we assume the equivalence also holds for $HFI_d$. There are two strong facts that support this. First, the equivalence between the two approaches demonstrates the mathematical equality of two parameters that already have the same physical meaning in each approach ($\lambda_g$ and $\Gamma_r$). Second, the fact that ECAE *R*(t) spectra present only one well-defined EFG (the final stable $EFG_f$), even in cases where there is no switch-off of the dynamic process, supports our scenario that the mean value $EFG_i$ of the initial fluctuating EFGs is the same as $EFG_f$.[16] At low temperatures, an initial EFG range (measured from $EFG_i$ = 200 MHz with a spread related to $\lambda_r$ = 39 MHz) correlates with a charge state of about  q = 0.55-, and at high temperatures with q = 0.95-, corresponding to $\lambda_r$= 8 MHz (see Figures 4 and 6, and the recent discussion). At RT, where dynamic effects are stronger, predictions become more challenging.  An enormous $\lambda_r$ value of 202 MHz spans all possible decreasing EFG values, indicating that at least five electron holes (q = 4+) are involved during the dynamic regime (see Figure 10). Figure 7a shows that this situation is energetically very unfavorable unless an external factor shifts the Fermi energy deep into the valence band. In our case, this extreme scenario arises from Auger cascades caused by $^{111}$In EC nuclear decay, followed by electronic relaxation and recombination at the $^{111}$Cd

atom, along with limited electron availability and/or mobility at this temperature in the current dilute semiconductor.

As described in detail, for $HFI_d$, we assigned the observed (fitted) *final* $EFG_f$ as a function of temperature to *stable* Cd`s charge states ranging from q=0.65- to q=1-. Only for intermediate temperatures, where the AE is stronger, and the number of "dynamic" electron holes exceeds one hole, these $EFG_f$ for $HFI_d$ could reflect a different *unstable* situation: a very fast fluctuation between two integer charge states, namely q=0 and q=1-. This strange effect was already detected in Cd-diffused $Lu_2O_3$ powder samples, where a single EFG varies with temperature following a Fermi-Dirac distribution.[39] Section SI.5 provides an explanation of why this scenario cannot originate the $EFG_f$ for $HFI_u$, nor for high and low temperatures for $HFI_d$.

Finally, we highlight the importance of establishing the equivalence between the two approaches. This allows us to identify $\lambda_r$ with the distribution half-width of the initial EFGs that produce the dynamic interaction and, therefore, determine the range of initial oscillating EFGs for each temperature and both hyperfine interactions using the BO model. Comparing these experimental quantities with *ab initio* predictions yields key insights into the electronic configurations of this complex doped system, both during and after the dynamic regime. On the other hand, the straightforward determination of the hole lifetime's temperature dependence using the BO approach in $^{111}$In-doped materials or other diluted systems that exhibit this *on-off* dynamic regime can be a powerful tool for studying conduction and transport properties in such systems.

## V. CONCLUSIONS

We demonstrated, through a comprehensive combined *ab initio*/DFT and double-approach experimental study, the conditions required to establish the equivalence of the perturbation factors

across very different methods proposed by Bäverstam *et al.* and Lupascu *et al.* for addressing dynamic *on-off* hyperfine interactions, such as those caused by the ECAE phenomenon.

For the diluted semiconductor studied here, ($^{111}$In→)$^{111}$Cd-doped α-$Al_2O_3$, these conditions were demonstrated for $HFI_u$, namely $EFG_i \approx EFG_f$. This equivalence enabled us to apply the BO approach to determine the range of initial EFGs generated by the dynamic electronic relaxation and recombination process, which, in this case, follows the electron capture that initiates the nuclear decay of $^{111}$In ions implanted in single crystals of α-$Al_2O_3$. This analysis was applied to both experimentally observed hyperfine interactions and at each temperature, also providing strong arguments for assuming the equivalence for $HFI_d$. If the L approach is used alone, it requires a series of TDPAC measurements at different temperatures to determine the initial EFGs, but it averages them over all temperatures. Of course, the final stable EFG reached for each interaction at each measured temperature is determined precisely by both approaches. Until now, the relaxation constant $\lambda_r$ of the BO model has been treated as a phenomenological parameter that describes the damping strength of TDPAC spectra caused by rapid, random EFG fluctuations during the dynamic phase of the ECAE phenomenon. However, this analysis found it to be half the width of the distribution of the initial fluctuating EFGs.

The mathematical equality found between the atomic recovery constant $\lambda_g$ and the relaxation rate $\Gamma_r$ within the equivalence, parameters that already have the same physical meaning, together with the excellent agreement between the experimentally determined ones for $HFI_u$ obtained independently by BO and L approaches, supports the equivalence proposed of both perturbation factors.

The very good match between the *ab initio* EFG predictions and the experimental final stable EFGs measured by the $^{111}$Cd probe for both interactions enabled us, through this integral analysis,

to connect the fluctuating initial EFGs observed experimentally with the various fluctuating electronic configurations that cause them.

We demonstrated that the experimental final EFG of $HFI_u$ corresponds to the charge state q = 1-, indicating that the impurity level is filled. We showed that the slight linear decrease in this EFG with increasing temperature results from significant nonlinear thermal lattice expansion, rather than from an electronic effect, thereby allowing us to distinguish between structural and electronic sources.

Conversely, the experimental *final* EFG of $HFI_d$ varies considerably with temperature. These EFGs correspond to the final stable electronic configurations the probe atom adopts at each temperature, influenced by the host material's electron availability and mobility. These configurations are correlated with charge states ranging from q = 1- to q = 0.65-. We showed that the significant increase in the EFG of $HFI_d$ with rising temperature results from the accumulation of more negative charge closer to the (001) plane than to the [001] crystalline axis at the Cd site, as electron holes ionization increases.

The agreement between the *final stable* EFGs reached by $HFI_d$ and the *initial* fluctuating EFGs for $HFI_u$ (determined independently using the L approach), together with our *ab initio* calculations of the EFG, supports the existence of different final electronic deficiencies trapped at the Cd impurity. These deficiencies are correlated with different charge states of the system, associated with a temperature-dependent Fermi energy. The Fermi energy fluctuates at each temperature during the dynamic regime, proportionally to the ECAE strength.

The application of the BO approach, enabling the overall analysis of $\tau_g$, now for *both* interactions and *each* measuring temperature, showed a good correlation with the variation of electron availability from intermediate to high temperatures, and with the availability and mobility of

electrons at very low temperatures, where the amplitude of the TDPAC signal is recovered. The behavior of $\tau_g$ enabled us to describe and quantify how the empty states (trapped electron holes) at $^{111}$Cd probes are filled during the dynamic process across the three selected temperature ranges. At high temperatures, the trapped electron holes when $\gamma_1$ is emitted live for about 5 ns, while at room temperature (RT), they live between 20 and 40 ns. At very low temperatures, the holes associated with $HFI_u$ live for 11 ns. In comparison, those associated with $HFI_d$ live for 45 ns because the initial fluctuating EFGs during the dynamic regime of this interaction involve more empty states. This is the maximum lifetime observed. Hence, it became clear that the reason dynamic hyperfine interactions were not observed in TDPAC experiments in $\alpha$-$Al_2O_3$ using $^{111m}$Cd as a probe is this short hole´s lifetime relative to the very long lifetime of the state emitting the $\gamma_1$-ray in this probe (48.6 min), not the absence of an EC, which would promote the formation of extra electron holes.

The range of the initial fluctuating EFGs for $HFI_u$ was determined using the BO (through $\lambda_r$ at each temperature) and L (through $\delta_i$ as an average) approaches, with good agreement between the two methods. These results were correlated with *ab initio* predictions, indicating the presence of 1 electron hole at RT for $HFI_u$. For $HFI_d$, only the BO approach was applied, indicating that at least 5 electron holes are trapped at the $^{111}$Cd atom at RT when the dynamic regime begins. Strictly speaking, the BO approach alone provides only the half-width of the initial fluctuating EFG distribution, not its range. However, within our proposed integrated scenario, the range was determined, noting that the initial EFGs´ distribution is centered at the $EFG_f$ value, when the equivalence is assumed.

We addressed the apparent contradiction that a spectrum related with a hyperfine dynamic process that never ends, as evidenced by continuous damping, displays a well-defined EFG, even

though a stable final $EFG_f$ does not exist in these cases. At least when the equivalence conditions are met, as $EFG_i = EFG_f$, it is $EFG_i$ that determines the well-defined shape of the spectrum. In this case, the perturbation factor for BO and L approaches takes the same form: a static $G_{22}$(t) damped by an exponential term. These findings also support our proposal that $EFG_f$ participates among the initial fluctuating EFGs, with the highest probability.

In summary, this detailed experimental and theoretical DFT study enabled us to identify, at each temperature, the initial fluctuating electronic configurations near the probe that give rise to the observed dynamic HFIs. It also established the final stable configurations, supporting and confirming the physical scenario underlying both the BO and L approaches used to model the ECAE phenomenon in diluted semiconductors and insulators. The methodology developed in this work can certainly be applied to future studies aimed at understanding transport and conduction properties in these materials.

## ASSOCIATED CONTENT

### Supporting Information

Details on a graphical road map of the BO and L approaches and the equivalence between them, the crystal structure and the Cd-doped cell used in the *ab initio* calculations, the parameters, parametrizations, and careful treatment for the use of fractional charge states, the defect formation energy calculations, and the two-state model discussion as a possible origin for the final $EFG_f$ are provided in an additional document (PDF).

## AUTHOR INFORMATION

### Corresponding Authors

*E-mail: darriba@fisica.unlp.edu.ar ; renteria@fisica.unlp.edu.ar

## ACKNOWLEDGEMENTS

CONICET and UNLP partially supported this work under Research Grants No. PIP0901 and No. 11/X1004, respectively, as well as by the Deutsche Forschungsgemeinschaft (DFG, German Research Foundation) under Grant No. VI 77/3Y1. This research used the computational facilities of the Physics of Impurities in Condensed Matter group at IFLP and the Department of Physics (UNLP). M.R. is grateful to Professors Dr. A.G. Bibiloni and Dra. C.P. Massolo for pioneering and encouraging discussions on the ECAE phenomenon, and G.N.D. and M.R. gratefully acknowledge Professor Dr. A.F. Pasquevich for fruitful discussions on the BO perturbation factor. Insightful talks about defect formation energy DFT calculations by Prof. Dra. L.V.C. Assali are kindly acknowledged. G.N.D. and M.R. are members of CONICET, Argentina.

## REFERENCES

(1) Dietl, T. A Ten-Year Perspective on Dilute Magnetic Semiconductors and Oxides. *Nat. Mater.* **2010**, *9*, 965–974. https://doi.org/10.1038/nmat2898.

(2) Green, R. J.; Regier, T. Z.; Leedahl, B.; McLeod, J. A.; Xu, X. H.; Chang, G. S.; Kurmaev, E. Z.; Moewes, A. Adjacent Fe-Vacancy Interactions as the Origin of Room Temperature

Ferromagnetism in $(In_{1-x}Fe_x)_2O_3$. *Phys. Rev. Lett.* **2015**, *115*, 167401. https://doi.org/10.1103/PhysRevLett.115.167401.

(3) Cai, X.; Sabino, F. P.; Janotti, A.; Wei, S.-H. Approach to Achieving a *p*-Type Transparent Conducting Oxide: Doping of Bismuth-Alloyed $Ga_2O_3$ with a Strongly Correlated Band Edge State. *Phys. Rev. B* **2021**, *103*, 115205. https://doi.org/10.1103/PhysRevB.103.115205.

(4) Zhi, J.; Zhou, M.; Zhang, Z.; Reiser, O.; Huang, F. Interstitial Boron-Doped Mesoporous Semiconductor Oxides for Ultratransparent Energy Storage. *Nat. Commun.* **2021**, *12, 445*. https://doi.org/10.1038/s41467-020-20352-4.

(5) Chaturvedi, S.; Waghmare, U. V. Origin of Ferromagnetism in a Two-Dimensional Dilute Magnetic Semiconductor h-$Co_xZn_{1-x}O$. *Phys. Rev. B* **2025**, *112*, 094437. https://doi.org/10.1103/2tcq-jqvw.

(6) Bibiloni, A. G.; Massolo, C. P.; Desimoni, J.; Mendoza-Zélis, L. A.; Sánchez, F. H.; Pasquevich, A. F.; Damonte, L.; López-García, A. R. Time-Differential Perturbed-Angular-Correlation Study of Pure and Sn-Doped $In_2O_3$ Semiconductors. *Phys. Rev. B* **1985**, *32*, 2393–2400. https://doi.org/10.1103/PhysRevB.32.2393.

(7) Massolo, C. P.; Desimoni, J.; Bibiloni, A. G.; Mendoza-Zélis, L. A.; Damonte, L. C.; Lopez-Garcia, A. R.; Martin, P. W.; Dong, S. R.; Hooley, J. G. Aftereffect Investigations in Mixed-Valence Indium Chlorides. *Phys. Rev. B* **1986**, *34*, 8857–8862. https://doi.org/10.1103/PhysRevB.34.8857.

(8) Bolse, W.; Uhrmacher, M.; Kesten, J. The EFG of the In-O-Bond $In_2O_3$, AgO and $Ag_2O$. *Hyperfine Interact.* **1987**, *35*, 931–934. https://doi.org/10.1007/BF02394526.

(9) Asai, K.; Ambe, F.; Ambe, S.; Okada, T.; Sekizawa, H. Time-Differential Perturbed-Angular-Correlation Study of Hyperfine Interactions at $^{111}$Cd($\leftarrow^{111}$In) in α-$Fe_2O_3$. *Phys. Rev. B* **1990**, *41*, 6124–6136. https://doi.org/10.1103/PhysRevB.41.6124.

(10) Bartos, A.; Lieb, K. P.; Pasquevich, A. F.; Uhrmacher, M. Scaling of the Electric Field Gradient of $^{111}$Cd Impurities in the Bixbyite Oxides of Y, Sc, Dy and Yb. *Phys. Lett. A* **1991**, *157*, 513–518. https://doi.org/10.1016/0375-9601(91)91029-D.

(11) Lupascu, D.; Habenicht, S.; Lieb, K.-P.; Neubauer, M.; Uhrmacher, M.; Wenzel, T. Relaxation of Electronic Defects in Pure and Doped $La_2O_3$ Observed by Perturbed Angular Correlations. *Phys. Rev. B* **1996**, *54*, 871–883. https://doi.org/10.1103/PhysRevB.54.871.

(12) Habenicht, S.; Lupascu, D.; Uhrmacher, M.; Ziegeler, L.; Lieb, K.-P. PAC-Studies of Sn-Doped $In_2O_3$: Electronic Defect Relaxation Following the $^{111}$In(EC)$^{111}$Cd-Decay. *Z. Phys. B* **1996**, *101*, 187–196. https://doi.org/10.1007/s002570050199.

(13) Habenicht, S.; Lupascu, D.; Neubauer, M.; Uhrmacher, M.; Lieb, K. P.; the ISOLDE-Collaboration. Doping of Sapphire Single Crystals with $^{111}$In and $^{111}$Cd Detected by Perturbed Angular Correlation. *Hyperfine Interact.* **1999**, *120*, 445–448. https://doi.org/10.1023/A:1017049523366.

(14) Penner, J.; Vianden, R. Temperature Dependence of the Quadrupole Interaction for $^{111}$In in Sapphire. *Hyperfine Interact.* **2004**, *158*, 389–394. https://doi.org/10.1007/s10751-005-9064-9.

(15) Darriba, G. N.; Muñoz, E. L.; Carbonari, A. W.; Rentería, M. Experimental TDPAC and Theoretical DFT Study of Structural, Electronic, and Hyperfine Properties in ($^{111}$In $\rightarrow$ )$^{111}$Cd-Doped $SnO_2$ Semiconductor: Ab Initio Modeling of the Electron-Capture-Decay After-Effects

Phenomenon. *J. Phys. Chem. C* **2018**, *122*, 17423–17436. https://doi.org/10.1021/acs.jpcc.8b03724.

(16) Darriba, G. N.; Muñoz, E. L.; Richard, D.; Ayala, A. P.; Carbonari, A. W.; Petrilli, H. M.; Rentería, M. Insights into the Aftereffects Phenomenon in Solids Based on DFT and Time-Differential Perturbed γ-γ Angular Correlation Studies in $^{111}$In(→$^{111}$Cd)-Doped Tin Oxides. *Phys. Rev. B* **2022**, *105*, 195201. https://doi.org/10.1103/PhysRevB.105.195201.

(17) Bäverstam, U.; Othaz, R.; de Sousa, N.; Ringström, B. After-Effects in the Decay of $^{75}$As and $^{197m}$Hg. *Nucl. Phys. A* **1972**, *186*, 500–512. https://doi.org/10.1016/0375-9474(72)90980-3.

(18) Iwatschenko-Borho, M.; Engel, W.; Foettinger, H.; Forkel, D.; Meyer, F.; Witthuhn, W. In-Beam Studies of Defect Cascades in Metals. *Nucl. Instrum. Methods Phys. Res. B* **1985**, *7–8*, 128–134. https://doi.org/10.1016/0168-583X(85)90542-7.

(19) Frauenfelder, H.; Steffen, R. M. *Alpha-, Beta-, and Gamma-Ray Spectroscopy*, K. Seigbahn.; North-Holland Publishing Co.: Amsterdam, Netherland, **1968**; Vol. 2.

(20) Kaufmann, E. N.; Vianden, R. J. The Electric Field Gradient in Noncubic Metals. *Rev. Mod. Phys.* **1979**, *51*, 161–214. https://doi.org/10.1103/RevModPhys.51.161.

(21) Schatz, G.; Weidinger, A. *Nuclear Condensed Matter Physics: Nuclear Methods and Applications*; Wiley: Chichester, England, **1996**.

(22) Mendoza-Zélis, L. A.; Bibiloni, A. G.; Caracoche, M. C.; Lopéz-García, A. R.; Martínez,J. A.; Mercader, R. C.;Pasquevich, A. F. Temperature dependence of the electric field gradient at Ta nuclei in hafnium pyrovanadate. *Hyperfine Interact.* **1977**, 3, 315-320. https://doi.org/10.1007/BF01021562.

(23) Darriba, G. N.; Faccio, R.; Eversheim, P.-D.; Rentería, M. Insights on the Relevance of DFT+U Formalism for Strongly Correlated Ta *d* Electrons Probing the Nanoscale in Oxides: Combined Time-Differential Perturbed γ-γ Angular Correlation Spectroscopy and *ab Initio* Study in $^{181}$Hf(→$^{181}$Ta)-Implanted α-$Al_2O_3$ Single Crystal. *Phys. Rev. B* **2023**, *108*, 245144. https://doi.org/10.1103/PhysRevB.108.245144

(24) Abragam, A.; Pound, R. V. Influence of Electric and Magnetic Fields on Angular Correlations. *Phys. Rev.* **1953**, *92*, 943–962. https://doi.org/10.1103/PhysRev.92.943.

(25) Pasquevich, A. F.; Rentería, M. Impurity Centers in Oxides Investigated by γ-γ Perturbed Angular Correlation Spectroscopy and Ab Initio Calculations. *Defect and Diffusion Forum* **2011**, *311*, 62–104. https://doi.org/10.4028/www.scientific.net/DDF.311.62.

(26) Winkler, H.; Gerdau, E. γγ-Angular Correlations Perturbed by Stochastic Fluctuating Fields. *Z. Physik* **1973**, *262*, 363–376. https://doi.org/10.1007/BF01394538.

(27) Winkler, H. γγ Angular Correlations Perturbed by Randomly Reorienting Hyperfine Fields. *Z Physik A* **1976**, *276*, 225–232. https://doi.org/10.1007/BF01412100.

(28) Uhrmacher, M.; Neubahuer, M.; Lupascu, D.; K. P. Lieb, K. P. *Proceedings of the International Workshop "25th Anniversary of Hyperfine interactions at La Plata"*, **1995** (*Departamento de Física, UNLP, La Plata, 1995)*, p. 82.

(29) Darriba, G. N.; Rentería, M.; Petrilli, H. M.; Assali, L. V. C. Site Localization of Cd Impurities in Sapphire. *Phys. Rev. B* **2012**, *86*, 075203. https://doi.org/10.1103/PhysRevB.86.075203.

(30) Madsen, G. K. H.; Blaha, P.; Schwarz, K.; Sjöstedt, E.; Nordström, L. Efficient Linearization of the Augmented Plane-Wave Method. *Phys. Rev. B* **2001**, *64*, 195134. https://doi.org/10.1103/PhysRevB.64.195134.

(31) Blaha, P.; Schwarz, K.; Madsen, G.; Kvasnicka, D.; Luitz, J. *WIEN2k, an Augmented Plane Wave Plus Local Orbitals Program for Calculating Crystal Properties*; Technical Universität: Wien, Austria, 2014.

(32) Herzog, P.; Freitag, K.; Reuschenbach, M.; Walitzki, H. Nuclear Orientation of $^{111m}$Cd in Zn and Be and the Quadrupole Moment of the 245keV State. *Z Physik A* **1980**, *294*, 13–15. https://doi.org/10.1007/BF01473117.

(33) Lucht, M.; Lerche, M.; Wille, H.-C.; Shvyd'ko, Y. V.; Rüter, H. D.; Gerdau, E.; Becker, P. Precise Measurement of the Lattice Parameters of α-Al2O3 in the Temperature Range 4.5–250 K Using the Mössbauer Wavelength Standard. *J. App. Crystallogr.* **2003**, *36*, 1075–1081. https://doi.org/10.1107/S0021889803011051.

(34) Ayala A. P. Program TDPAC, Code Developed to Fit Static and Dynamic Perturbation Factors for Multiple Sites for Polycrystalline Samples, PhD Thesis, Universidad Nacional de La Plata, 1995.

(35) Lohstroh, A.; Uhrmacher, M.; Wilbrandt, P.-J.; Wulff, H.; Ziegeler, L.; Lieb, K. P. Electronic Relaxation in Indium Oxide Films Studied with Perturbed Angular Correlations. *Hyperfine Interact.* **2004**, *159*, 35–42. https://doi.org/10.1007/s10751-005-9078-3.

(36) Muñoz, E. L.; Richard, D.; Carbonari, A. W.; Errico, L. A.; Rentería, M. PAC Study of Dynamic Hyperfine Interactions at $^{111}$In-Doped $Sc_2O_3$ Semiconductor and Comparison with Ab

Initio Calculations. *Hyperfine Interact.* **2010**, *197*, 199–205. https://doi.org/10.1007/s10751-010-0207-2.

(37) E. L. Muñoz. Estudio experimental y de primeros principios de interacciones hiperfinas dinámicas en óxidos semiconductores dopados con impurezas ($^{111}$In(EC)→)$^{111}$Cd. PhD Thesis, Universidad Nacional de La Plata, 2011. https://sedici.unlp.edu.ar/handle/10915/2628

(38) Kehr, K. W. Empirical Information on Quantum Diffusion. *Hyperfine Interact.* **1984**, *17*, 63–74. https://doi.org/10.1007/BF02065887.

(39) Errico, L. A.; Rentería, M.; Bibiloni, A. G.; Darriba, G. N. Temperature dependence of the EFG at Cd-doped $Lu_2O_3$: How *ab initio* calculations can complement PAC experiments. (2005), Phys. Stat. Sol. (c) **2005**, 2, 3576-3580. https://doi.org/10.1002/pssc.200461788.

TOC Graphic

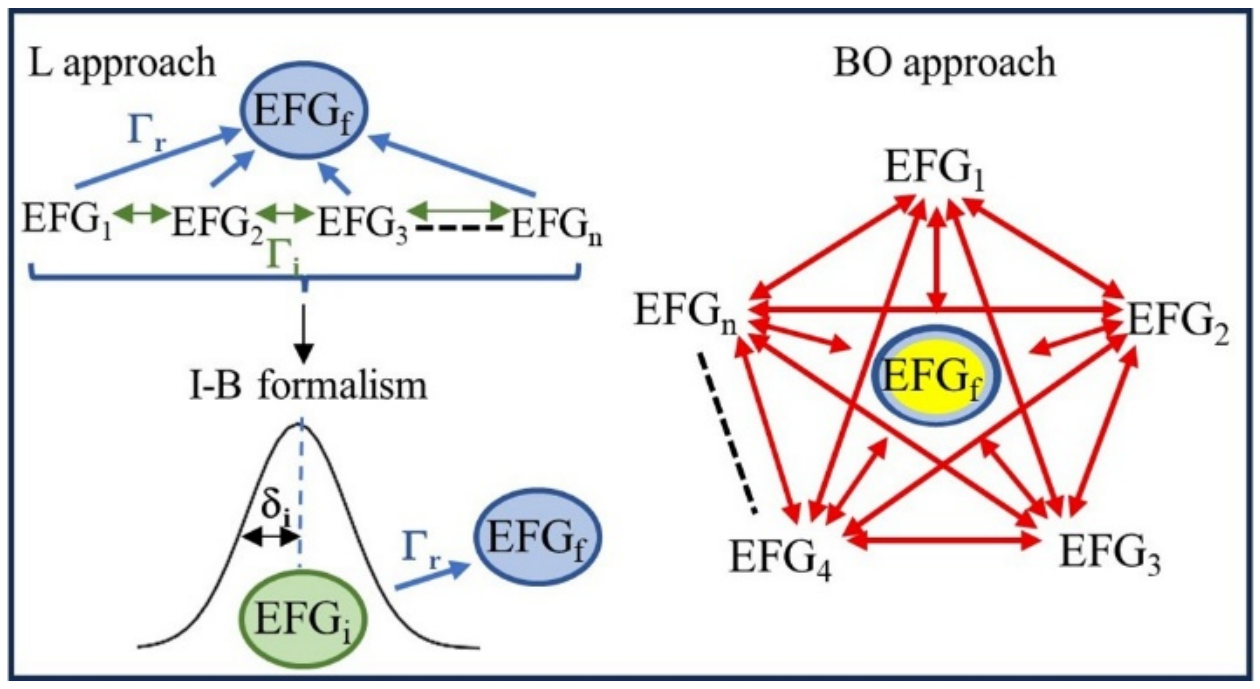

# SUPPORTING INFORMATION

## “Aftereffects” Phenomenon in $^{111}In(\rightarrow^{111}Cd)$-Implanted $\alpha$-$Al_2O_3$ Single Crystals: Novel Approach Integrating Experimental Double-Model Analysis with Density-Functional Theory

*Germán N. Darriba* [*,†], *Reiner Vianden* [‡], *Alejandro P. Ayala* [§], *and Mario Rentería* [*,†]

† Departamento de Física “Prof. Dr. Emil Bose” and Instituto de Física La Plata [IFLP, Consejo Nacional de Investigaciones Científicas y Técnicas (CONICET) La Plata], Facultad de Ciencias Exactas, Universidad Nacional de La Plata, CC 67, (1900) La Plata, Argentina.

‡ Helmholtz-Institut für Strahlen- und Kernphysik (HISKP), Universität Bonn, Nussallee 14–16, 53115 Bonn, Germany

§ Departamento de Física, Universidade Federal do Ceará, Fortaleza, CE, 60644-900, Brazil.

**Corresponding Authors**: darriba@fisica.unlp.edu.ar ; renteria@fisica.unlp.edu.ar

**Section SI.1: Summary of the experimental integrated double-model approach**

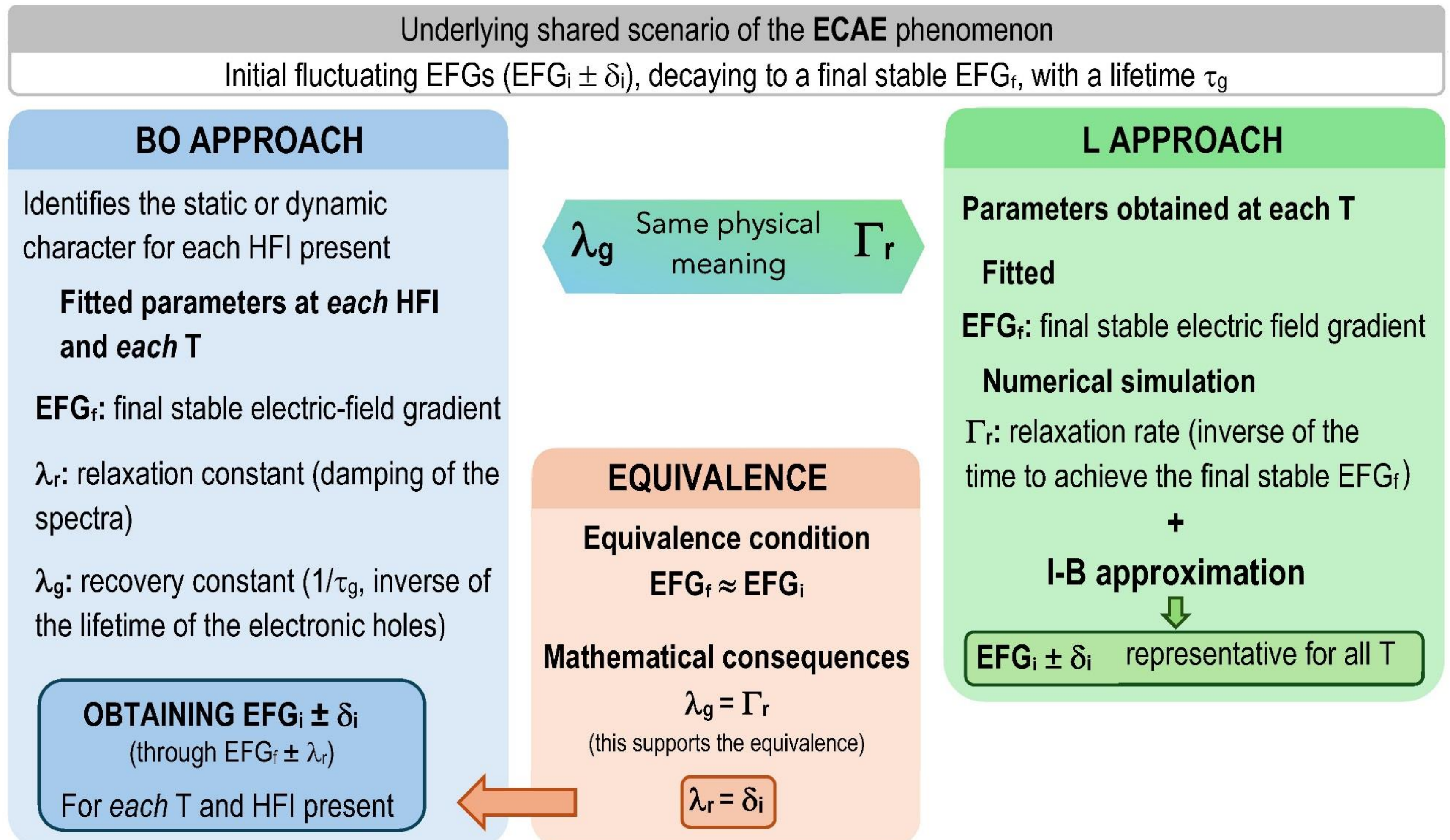

**Figure SI.1**: Schematic representation of the integrated double-model approach. $\delta_i$ is the half-width of the distribution of initial fluctuating EFGs ($EFG_i$) in the I-B formalism.

**Section SI.2: Crystal structure and the Cd-doped cell used in the calculations**

We use the hexagonal representation of its unit cell, which contains three rhombohedral primitive cells (space group $R\overline{3}c$), as shown in Figure SI.2. In this representation, the structure has $a = b =$ 4.75999(3) Å and $c =$ 12.99481(7) Å,[1] with 12 Al and 18 O atoms at ±(0, 0, $u$); ±(0, 0, $u$+1⁄2); *rh* and ±($v$, 0, 1⁄4); ±(0, $v$, 1⁄4); ±(−$v$, −$v$, 1⁄4); *rh* positions, respectively, where $u$ = 0.35219(1) and $v$ = 0.30633(5). The "*rh*" term implies adding (1⁄3, 2⁄3, 2⁄3) and (2⁄3, 1⁄3, 1⁄3) to the previous coordinates. In this oxide, each Al atom has six nearest oxygen neighbors (ONN), three (O1) at 1.854 Å and three (O2) at 1.972 Å.

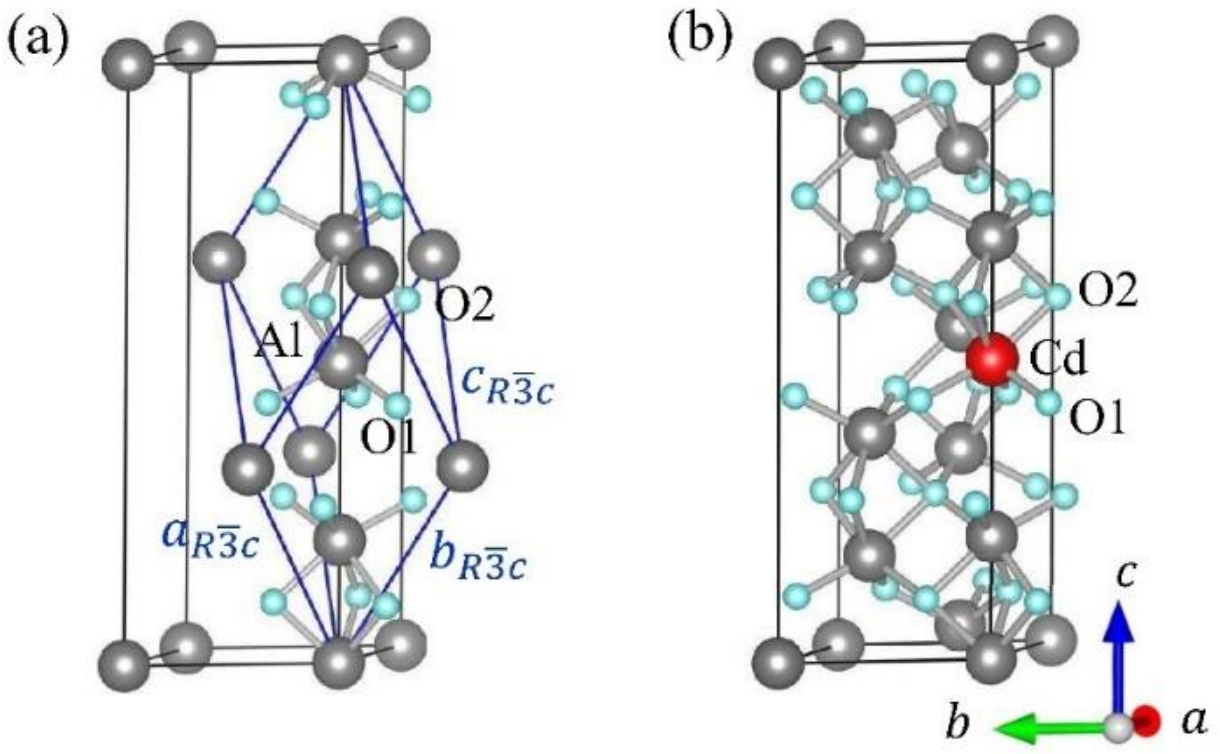

**Figure SI.2:** (a) Rhombohedral α-$Al_2O_3$ primitive cell. Gray and light blue spheres indicate Al and O atoms, respectively. (b) Cd-doped α-$Al_2O_3$ hexagonal unit cell used in all calculations, with one Al atom replaced by a Cd impurity (red sphere).

To simulate the condition of diluted impurity, where each impurity does not interact with its nearest images and the structural relaxation caused by the impurity in its neighbors does not

influence the relaxations induced by the closest impurities, we replace one of the 12 Al atoms with a Cd atom (Figure SI.2b), as shown to be sufficient for this doped system.[2]

This substitution induces significant forces on its nearest-neighbor oxygen atoms. Therefore, to find the equilibrium structure, all atomic displacements in the cell were allowed in our calculations. As shown in ref. 2, the presence of the Cd impurity results in outward relaxations of O1 and O2, increasing the bond lengths dCd-O1 and dCd-O2 by 11% and 14 %, respectively.

On the other hand, to account for thermal effects on the structural parameters of $\alpha$-$Al_2O_3$, we use experimental lattice parameters as a function of temperature.[3]

## Section SI.3: Parameters and parametrizations used in the *ab initio* calculations

For all APW+lo calculations, we set a cutoff parameter for the plane-wave basis $R_{MT}K_{max}$ = 7, where $K_{max}$ is the largest magnitude of the reciprocal lattice vectors, and $R_{MT}$ is the smallest radius of the non-overlapping muffin-tin spheres. The radii of the spheres centered on the Cd, Al, and O atoms were $R_{MT}$(Cd) = 1.06 Å, $R_{MT}$(Al) = 0.87 Å, and $R_{MT}$(O) = 0.85 Å, respectively. The integration in reciprocal space was performed using the tetrahedron method,[4] with a *k*-space grid of 9x9x2. The Perdew-Burke-Ernzerhof (PBE-GGA) parametrization[5] was employed to treat exchange-correlation. We also used the LDA functional for checking purposes, obtaining the same qualitative results and the same EFG predictions for the relaxed PBE-GGA positions. Finally, to determine the equilibrium structures in all cases, a Newton-damped scheme was used until the forces on the ions dropped below 0.01 eV/Å, and the EFG tensors were calculated from the second derivative of the full electric potential obtained for the final equilibrium structures.[6,7]

The use of fractional charge states in these DFT calculations requires careful treatment and checking. In certain systems, when using common functionals, fractional charges usually violate the physical condition that energy should behave linearly between integer charge numbers, may produce excessive charge delocalization, and/or self-interaction errors. In our case, for a semiconductor or wide-band-gap insulator, energy behaves linearly between integer charge numbers and, more importantly for this study, the EFG (which is more sensitive than energy) behaves smoothly and continuously between the EFG values predicted for integer charge states, namely from q=0 to q=1-. In addition, most of the added charge is highly localized at the Cd atom, and the rest is localized at the nearest oxygen neighbors. This is valid for 1 added electron (as shown in Figure 3 of the manuscript), and also when fractions of it are added to the system. Finally, these added charges are localized in energy, as shown in the DOS of Figure 2 of the manuscript.

**Section SI.4: Defect-formation energy study of a Cd impurity in α-$Al_2O_3$ as a function of the system´s charge state**

For the Cd-doped α-$Al_2O_3$ system, the defect formation energy in a charge state $q$ is:

$$E_f(\mathrm{Al_2O_3{:}Cd})^q = E^q(n_{Al}, n_O, n_{Cd}) - n_{Al}\mu_{Al} - n_O\mu_O - n_{Cd}\mu_{Cd} + q(\varepsilon_F + \varepsilon_v').$$

Here, $E^q(n_{Al}, n_O, n_{Cd})$ is the total energy of the Cd-doped α-$Al_2O_3$ system in the charge state $q$, with $n_{Al}$ , $n_O$, and $n_{Cd}$ atoms of aluminum, oxygen, and cadmium, being their chemical potentials

$\mu_{Al}$, $\mu_{O}$, and $\mu_{Cd}$, respectively. $\varepsilon_F$ is the Fermi energy, relative to the energy of the top of the valence band of the doped system, $\varepsilon_v'$ . $\varepsilon_F$ takes values between 0 and the band gap energy $\varepsilon_g$ ($0 \leq \varepsilon_F \leq \varepsilon_g$). Following the formalism developed in ref. 2, we have:

$$E_f(\mathrm{Al_2O_3{:}Cd})^q = E(\mathrm{Al_2O_3{:}Cd})^q - E(\mathrm{Al_2O_3}) + \mu_{Al}^* + \frac{1}{2}\Delta_f H^{\mathrm{Al_2O_3}} - \mu_{Cd}^* - \Delta_f H^{\mathrm{CdO}} + q(\varepsilon_F + \varepsilon_v')$$

where $\mu_x^*$ is the total energy per atom of the metallic crystal *x*, and $\Delta_f H^{\mathrm{Y}}$ is the formation enthalpy of the compound Y.

**Section SI.5: Two-state model discussion as a possible origin for the final $EFG_f$**

The fitted (observed) final EFGs, $EFG_f$, can reflect a constant-in-time electronic environment, as usually occurs, or in turn can arise from, e.g., a very fast fluctuation between two probe´s charge states or electronic configurations. This is rarer, but we have observed this effect, e.g., in TDPAC experiments on powder samples of $^{111}$In(EC→$^{111}$Cd)-diffused $Lu_2O_3$.[8] In these cases, the determined single EFG, variable and reversible with measuring temperature, results from the weighted average of the EFGs of each probe´s state: EFG (T) = A(T) EFG1 + B(T) EFG2, with A+B = 1, and EFG1 and EFG2 the EFG produced by charge state 1 and state 2, respectively. The fractions A and B are proportional to the time the probe lives in each quantum state and vary with temperature T. When this occurs, the EFG varies with temperature following a Fermi-Dirac distribution.

The final $EFG_f$ of $HFI_u$ is the upper possible value in this material (corresponding to adding 1 electron, q=1-); it has no larger values above it (since adding another electron is not physical), and hence it cannot be an average of two integer values.

The case of $HFIf_d$ is different. At any temperature, their final EFGs have values both lower and higher, hence these $EFG_f$ can be the result of a fast fluctuation between two EFGs related to integer charge states, namely $Cd^0$ (q=0, 1 electron hole) and $Cd^{1-}$ (q=1-, impurity level completely filled).

But at both high and low temperatures, during the dynamic process, the number of trapped electron holes is much less than 1, i.e., the impurity level is already almost filled before the dynamic interaction ends. So, it makes no sense that, after the switch-off of the dynamic process, the system goes back (to the $Cd^0$ state) and fluctuates between the $Cd^0$ and $Cd^{1-}$ states. Also, the $EFG_f$ in this case is so close to that of $Cd^{1-}$ that the probe would be in the $Cd^{1-}$ state almost all the time.

Conversely, at intermediate temperatures, e.g., at 300 K, we know that for $HFI_d$ there are at least 5 electron holes during the dynamic process; hence, the final $EFG_f$ in this case (related nominally to q=0.65-) could result from a very fast fluctuation between $Cd^0$ and $Cd^{1-}$ states. The same could occur at proximal increasing temperatures, provided the number of dynamic holes always exceeds 1.

In summary, we conclude that only for $HFI_d$ and when the AE is strong can the two-state scenario be possible, and, provided that the EFG distribution parameter is very low for the two interactions across the entire measurement temperature range, we assign the $EFG_f$ to stable final charge-state configurations.

## References

(1) Izumi, F.; Asano, H.; Murata, H.; Watanabe, N. Rietveld Analysis of Powder Patterns Obtained by TOF Neutron Diffraction Using Cold Neutron Sources. *J. App. Crystallogr.* **1987**, *20*, 411–418. https://doi.org/10.1107/S0021889887086382.

(2) Darriba, G. N.; Rentería, M.; Petrilli, H. M.; Assali, L. V. C. Site Localization of Cd Impurities in Sapphire. *Phys. Rev. B* **2012**, *86*, 075203.

https://doi.org/10.1103/PhysRevB.86.075203.

(3) Lucht, M.; Lerche, M.; Wille, H.-C.; Shvyd'ko, Y. V.; Rüter, H. D.; Gerdau, E.; Becker, P. Precise Measurement of the Lattice Parameters of α-Al2O3 in the Temperature Range 4.5–250 K Using the Mössbauer Wavelength Standard. *J. App. Crystallogr.* **2003**, *36*, 1075–1081. https://doi.org/10.1107/S0021889803011051.

(4) Blöchl, P. E. Projector Augmented-Wave Method. *Phys. Rev. B* **1994**, *50*, 17953–17979. https://doi.org/10.1103/PhysRevB.50.17953.

(5) Perdew, J. P.; Burke, K.; Ernzerhof, M. Generalized Gradient Approximation Made Simple. *Phys. Rev. Lett.* **1996**, *77*, 3865–3868. https://doi.org/10.1103/PhysRevLett.77.3865.

(6) Schwarz, K.; Ambrosch-Draxl, C.; Blaha, P. Charge Distribution and Electric-Field Gradients in $YBa_2Cu_3O_{7-x}$. *Phys. Rev. B* **1990**, *42*, 2051–2061.

https://doi.org/10.1103/PhysRevB.42.2051.

(7) Blaha, P.; Schwarz, K.; Dederichs, P. H. First-Principles Calculation of the Electric-Field Gradient in Hcp Metals. *Phys. Rev. B* **1988**, *37*, 2792–2796.

https://doi.org/10.1103/PhysRevB.37.2792.

(8) Errico, L. A.; Rentería, M.; Bibiloni, A. G.; Darriba, G. N. Temperature dependence of the EFG at Cd-doped $Lu_2O_3$: How *ab initio* calculations can complement PAC experiments. (2005), Phys. Stat. Sol. (c) **2005**, 2, 3576-3580. https://doi.org/10.1002/pssc.200461788